\documentclass[nofootinbib, twocolumn,superscriptaddress,preprintnumbers,letterpaper,nofootinbib,natbib]{revtex4-1}
\pdfoutput = 1
\usepackage{amssymb}
\usepackage{amsmath}
\usepackage{epsfig}
\usepackage[usenames,dvipsnames]{xcolor}
\usepackage[breaklinks,colorlinks,urlcolor=blue,citecolor=blue,linkcolor=blue]{hyperref}
\usepackage{cleveref}
\usepackage{multirow}
\usepackage{capt-of}
\usepackage{siunitx}
\usepackage{graphicx}
\usepackage{slashed}
\usepackage{tabularx}
\usepackage{multirow}
\usepackage{braket}
\usepackage{amsfonts}
\usepackage{mathtools}
\usepackage{mathrsfs}
\usepackage[compat=1.1.0]{tikz-feynman}
\usepackage{subfiles}
\usepackage{slashed}
\usepackage{comment}

\usetikzlibrary{positioning}
\usepackage[compatibility=false]{caption}
\usepackage{subcaption}

\usepackage[normalem]{ulem}
\usepackage[shortlabels]{enumitem}  

\usepackage{fontawesome}

\setcitestyle{sort&compress}

\usepackage{float}

\newcommand{\op}[2]{ \mathcal{O}_{#1} ^{#2} } 

\newcommand{\coeff}[2]{ \mathcal{C}_{#1} ^{#2} }
\DeclareFontFamily{OT1}{pzc}{}
\DeclareFontShape{OT1}{pzc}{m}{it}{<-> s * [1.10] pzcmi7t}{}
\DeclareMathAlphabet{\mathpzc}{OT1}{pzc}{m}{it}
\newcommand{\adcoeff}[2]{ \mathpzc{c}_{#1} ^{#2} } 
\newcommand{\highscale}{\Lambda}
\newcommand{\opdim}{\mathfrak{D}}

\newcommand{\group}[1]{\mathrm{#1}}
 
\newcommand{\adm}{ \Gamma}
\newcommand{\hc}{\mathrm{H.c.}}

\newcommand{\yuk}[1]{{Y}_{#1}}

\newcommand{\mtrx}[1]{\mathcal{M}_{#1}}

\newcommand{\order}[1] {\mathcal{O }\left( #1 \right)}

\newcommand{\lag}{ \mathcal{L}} 

\makeatletter
\newcommand{\overleftrightsmallarrow}{\mathpalette{\overarrowsmall@\leftrightarrowfill@}}
\newcommand{\overrightsmallarrow}{\mathpalette{\overarrowsmall@\rightarrowfill@}}
\newcommand{\overleftsmallarrow}{\mathpalette{\overarrowsmall@\leftarrowfill@}}
\newcommand{\overarrowsmall@}[3]{%
	\vbox{%
		\ialign{%
			##\crcr
			#1{\smaller@style{#2}}\crcr
			\noalign{\nointerlineskip}%
			$\m@th\hfil#2#3\hfil$\crcr
		}%
	}%
}
\def\smaller@style#1{%
	\ifx#1\displaystyle\scriptstyle\else
	\ifx#1\textstyle\scriptstyle\else
	\scriptscriptstyle
	\fi
	\fi
}
\makeatother

\newcommand{\refsec}[1]{{Sec.~\ref{#1}} }

\pgfmathsetmacro\sizedot{1.1}
\pgfmathsetmacro\sizesqdot{1.5}
\pgfmathsetmacro\sizecrodot{1.0}

\makeatletter

\begin{document}

\title{
Renormalisation group running effects in $pp\to t\bar{t}h$ in the Standard Model Effective Field Theory
}

\author{Stefano Di Noi}
\email{stefano.dinoi@phd.unipd.it}

\author{Ramona Gr\"ober}
\email{ramona.groeber@pd.infn.it}
\affiliation{Dipartimento di Fisica e Astronomia "Galileo Galilei", Universit\`a di Padova, Italy and Istituto Nazionale di Fisica Nucleare, Sezione di Padova, Padova, I-35131, Italy}

\begin{abstract} 
\centering 
We study the effects of renormalisation group running of the Wilson coefficients in Standard Model Effective Field Theory, using the process $pp \to t \bar{t}h$ as a showcase. We consider both strong and top Yukawa running effects, since the latter can be relevant in presence of large Wilson coefficients. We study the difference between the use of a dynamical and fixed renormalisation scale by exploring different scenarios for the higher-dimensional operators at the high energy scale of new physics, assumed to be at the TeV scale. 
\end{abstract}

\maketitle

\tableofcontents

\setlength\parskip{4pt}

\section{Introduction}\label{sec:intro}
After the discovery of a scalar particle compatible with the properties of the Standard Model (SM) Higgs boson in 2012 \cite{HiggsATLAS,HiggsCMS}, all the particles predicted by the SM have been observed. 
However, various reasons suggest that the SM should be extended, as for instance massive neutrinos, the matter anti-matter asymmetry of the universe or the missing dark matter candidate. \par 
Given that, so far, no striking new physics signal has been observed the new physics energy scale might lie much above the electroweak scale. This motivates to use an effective field theory approach to describe new physics in a model-independent way. 
In this context, the Standard Model Effective Field Theory (SMEFT) \cite{Buchmuller:1985jz,dim6smeft} represents a powerful tool to approach the search for heavy new physics. \par
Let $\highscale \gtrsim v=246\, \mathrm{GeV}$ be the typical mass scale of the new degrees of freedom, lying beyond the experimental reach. The indirect effects of such heavy particles can be captured and parametrised with a tower of higher dimensional operators:
\begin{equation}
\label{eq:SMEFT1}
\lag_{\mathrm{SMEFT}} =  \lag_{\mathrm{SM}} + \sum_{\opdim_i > 4} \coeff{i}{} \op{i}{},
\end{equation}
where $\op{i}{}$ is an operator consisting of SM fields (and derivatives) invariant under the SM unbroken gauge group $\group{SU(3)}_{\mathrm{C}} \otimes\group{SU(2)}_{\mathrm{L}}\otimes\group{U(1)}_{\mathrm{Y}}$. 
Assuming lepton and baryon number conservation, the first terms in Eq.~\eqref{eq:SMEFT1} arise at dimension-six level and are expected to give the dominant contributions in collider physics.  A complete and non-redundant basis is defined in Ref.~\cite{dim6smeft} (\textit{Warsaw basis}) and in Refs.~\cite{Giudice:2007fh, Contino:2013kra, Elias-Miro:2013eta} (\textit{SILH basis}). 

Loop computations typically give an infinite result: to get rid of such divergences, the theory must be renormalised.\footnote{Formally, the Lagrangian in Eq.~\eqref{eq:SMEFT1} is not renormalisable since it contains terms with mass dimension greater than four. This implies that the absorption of divergences requires an infinite number of counterterms. However, is it possible to renormalise the theory order by order in the expansion in $1/\Lambda$, using a finite number of counterterms, see e.g.~Ref.~\cite{man} for a more detailed discussion.} An important consequence is that the parameters in the theory acquire a dependence on the energy scale, encoded in a set of differential equations known as Renormalisation Group Equations (RGEs). 
For what concerns the SMEFT at dimension-six and one-loop level, the RGEs have been computed in Refs.~\cite{rge1,rge2,rge3}.

Typically, the SMEFT is used as the low-energy limit of some UV model (top-down approach). At some energy scale $\mu \sim \highscale$, where $\highscale$ may be the typical mass scale of the new degrees of freedom, a matching between the UV model and the SMEFT is performed.
The tree-level matching between the SMEFT and generic UV theories has been computed and automatised and one-loop matching is in development (see Refs.~\cite{Criado:2017khh,DasBakshi:2018vni, Cohen:2020qvb, Fuentes-Martin:2022jrf,Guedes:2023azv}). After the matching, the RGE running can be computed in the SMEFT, making it possible to address the impact of the higher dimensional operators on low-energy observables such as cross sections and differential distributions. 
This program offers a universal and pragmatic approach to study the phenomenological consequences of new physics including running effects that would otherwise need an \textit{ad hoc} computation for each UV model. 
In the bottom-up approach instead one is agnostic on the Wilson coefficients of the higher-dimensional operators and uses them to parametrise deviations with respect to the SM in a model-independent way. In this way, the bounds on the coefficients can be used to restrict the parameter space of any new physics scenario.

The goal of this paper is to test the effects of RGE running in the prediction of differential cross sections when adopting a dynamical scale choice. This means that rather than evolving the SMEFT RGEs down to some fixed electroweak scale we evolve them to a dynamical scale varying event by event. We will showcase this effect by considering the example of the process $pp\to t\bar{t}h$. The running effects proportional to $\alpha_s$ have been recently studied in Refs.~\cite{Maltoni:2016yxb,Aoude:2022aro,Grazzini:2018eyk, Battaglia:2021nys}. In this work, we take a step further and consider the fully coupled system of the SMEFT RGEs. The motivation is that, in particular for operators in the top sector, the running is often proportional to the top-quark Yukawa coupling. Such operators tend to be less constrained than operators whose running is proportional to $\alpha_s$ (see Ref.~\cite{Ethier:2021bye}). For this reason, these effects can be hence of similar size or even bigger as the $\alpha_s$ ones. In addition, we show that the first leading-log approximation, in which the RGEs are solved assuming that the anomalous dimension matrix is independent of the scale, is not sufficient to describe accurately the RGE running effects in the presence of large Wilson coefficients. This motivates to solve numerically the fully coupled system of RGEs in collider physics studies.

The paper is organized as follows: in \refsec{sec:computation} we describe the relevant set of operators and our computational strategy.
In \refsec{sec:results} we show the numerical results of our analysis in three different scenarios for the Wilson coefficients.
Finally, in \refsec{sec:conclusions} we give our conclusions.

\section{Description of the computation} \label{sec:computation}
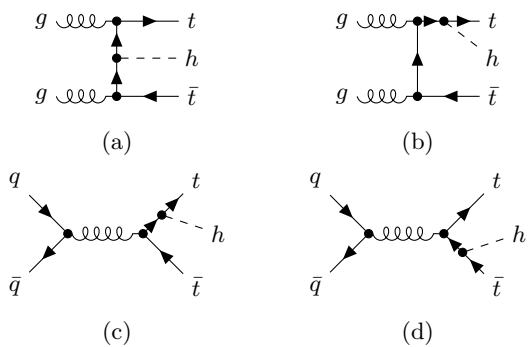
\begin{figure}[h]
    \centering
    \begin{subfigure}[t]{0.45\linewidth}
        \centering
        \begin{tikzpicture}[baseline=(h)]
            \begin{feynman}[small]
                \vertex  (gi1) {\(  g \)};
                \vertex (gi2) [below = of gi1] {\( g \)};
                \vertex (gtt1) [dot, scale = \sizedot, right = of gi1] {};
                \vertex (gtt2) [dot, scale = \sizedot, right = of gi2] {};
                \vertex (htt) [dot, scale=\sizedot, below =14pt of gtt1] {};
                \vertex (h) [right = of htt] {\( h \)};
                \vertex (tf1) [right = of gtt1] {\( t \)};
                \vertex (tf2) [right = of gtt2] {\( \bar{t} \)};
                \diagram* {
                    (gi1)  -- [gluon] (gtt1),
                    (gi2)  -- [gluon] (gtt2),
                    (tf2)  -- [fermion] (gtt2) -- [fermion] (htt) -- [fermion] (gtt1) -- [fermion] (tf1),
                    (h) -- [scalar] (htt)
                };
            \end{feynman}
        \end{tikzpicture}
        \caption{}\label{fig:ggtthSM1}
    \end{subfigure}
      \begin{subfigure}[t]{0.45\linewidth}
        \centering
        \begin{tikzpicture}[baseline=(h)]
			\begin{feynman}[small]
				\vertex  (gi1) {\(  g \)};
				\vertex (gi2) [below = of gi1] {\( g \)};
				\vertex (gtt1) [dot, scale = \sizedot, right = of gi1] {};
				\vertex (gtt2) [dot, scale = \sizedot, right = of gi2] {};
				\vertex (htt) [dot, scale=\sizedot, right=10pt of gtt1] {};
				\vertex (h) [below = 14 pt of tf1] {\( h \)};
				\vertex (tf1) [right = of gtt1] {\( t \)};
				\vertex (tf2) [right = of gtt2] {\( \bar{t} \)};
				\diagram* {
					(gi1)  -- [gluon] (gtt1),
					(gi2)  -- [gluon] (gtt2),
					(tf2)  -- [fermion] (gtt2)  -- [fermion] (gtt1) -- [fermion] (htt)-- [fermion] (tf1),
					(h) -- [scalar] (htt)
				};
		\end{feynman}  \end{tikzpicture}
        \caption{}\label{fig:ggtthSM2}
    \end{subfigure}
    \begin{subfigure}[t]{0.45\linewidth}
        \centering
        \begin{tikzpicture}[baseline=(h)]
            \begin{feynman}[small]
                \vertex (q1) {\(  q \)};
                \vertex (qqg1) [dot, scale=\sizedot, below right=of q1] {};
                 \vertex (q2) [below left =  of qqg1] {\( \bar{q} \)};
                \vertex (qqg2) [dot, scale=\sizedot,right = of qqg1] {};
                \vertex (h) [right = of qqg2] {\( h \)};                
                \vertex (tth) [dot, scale=\sizedot,above right = 10 pt of qqg2] {};
                \vertex  (t1) [above right= of qqg2]{\(  t \)};
                \vertex  (t2) [below right= of qqg2]{\(  \bar{t} \)};
                \diagram* {
                    (q1) --[fermion] (qqg1) -- [fermion] (q2),
                    (qqg1) --[gluon] (qqg2),
                    (t2) -- [fermion] (qqg2) --[fermion] (tth) -- [fermion] (t1),
                    (h) -- [scalar] (tth)
                };
            \end{feynman}
        \end{tikzpicture}
        \caption{}\label{fig:qqtthSM1}
    \end{subfigure}
      \begin{subfigure}[t]{0.45\linewidth}
        \centering
        \begin{tikzpicture}[baseline=(h)]
            \begin{feynman}[small]
                \vertex (q1) {\(  q \)};
                \vertex (qqg1) [dot, scale=\sizedot, below right=of q1] {};
                 \vertex (q2) [below left =  of qqg1] {\( \bar{q} \)};
                \vertex (qqg2) [dot, scale=\sizedot,right = of qqg1] {};
                \vertex (h) [right= of qqg2] {\( h \)};                
                \vertex (tth) [dot, scale=\sizedot,below right = 10 pt of qqg2] {};
                \vertex  (t1) [above right= of qqg2]{\(  t \)};
                \vertex  (t2) [below right= of qqg2]{\(  \bar{t} \)};
                \diagram* {
                    (q1) --[fermion] (qqg1) -- [fermion] (q2),
                    (qqg1) --[gluon] (qqg2),
                    (t2) --[fermion] (tth) -- [fermion] (qqg2)  -- [fermion] (t1),
                    (h) -- [scalar] (tth)
                };
            \end{feynman}
        \end{tikzpicture}
        \caption{}\label{fig:qqtthSM2}
    \end{subfigure}
\caption{Tree-level diagrams for $pp\to t\bar{t}h$ in the SM.} \label{fig:pptthSM}
\end{figure}

The hadronic process $pp\to t \bar{t} h$ is generated by two different partonic channels: $gg\to t\bar{t} h$ and $\bar{q}q \to t\bar{t} h$, being $q=u,d,c,s,b$.
In the SM this process arises at tree-level via diagrams like the ones in Fig.~\ref{fig:pptthSM}.

Considering in addition SMEFT operators gives rise to new interaction vertices, as well as to modifications of the existent ones. The complete set of Feynman rules for the SMEFT (in the broken phase), which we follow here, is given in Ref.~\cite{FeynRules}.

The Lagrangian considered in this work is:
\begin{equation}
\lag_{t\bar{t}H}  =  \lag_{\mathrm{SM}} + \lag_{\phi} + \lag_{\mathrm{B}} + \lag_{\mathrm{4F}} .
\label{eq:Operators}
\end{equation}

The SM Lagrangian in Eq.~\eqref{eq:Operators} is given by
\begin{equation}
\label{eq:SMlag}
\begin{split}
\lag_{\textrm{SM}} = & -\frac{1}{4} G_{\mu \nu}^A G^{A\mu \nu}
+ \sum_{\psi} \bar{\psi} i \slashed{D} \psi 
+ (D_{\mu} \phi^\dagger) (D^{\mu} \phi) \\ 
&- \lambda \left(\phi^\dagger \phi - \frac{1}{2}v^2 \right)^2 
- \left[ 
\tilde{\phi}^\dagger \bar{t}_R\, \yuk{t}\, Q_L
+ \hc   \right],
\end{split}
\end{equation}
where $Q_L$ and $t_R$ refer to the
third family isospin doublet (left-handed) and isospin singlet (right-handed, up-type), respectively, $\tilde{\phi_j} \;  =\;\varepsilon_{jk} \phi^{\dagger k}$ (with $\varepsilon_{12}=+1$) and $T^A$ are the $\group{SU(3)}_\mathrm{C}$ generators. 
When spontaneous symmetry breaking occurs, the physical Higgs boson acquires a vacuum expectation value $v = 246$ GeV leading to a mass given by $m^2_H = 2 \lambda v^2$ (in the SM). 

The part of the SMEFT Langrangian containing only the Higgs doublet and its gauge interactions is given by
\begin{equation}\begin{split}\label{eq:Lphi}
\lag_{\phi} = & \coeff{\phi}{} \left( \phi^\dagger \phi  \right)^3 + \coeff{\phi \Box}{}  \left( \phi^\dagger \phi  \right) \Box \left( \phi^\dagger \phi  \right)  \\ + & \coeff{\phi D}{} \left( \phi^\dagger D_{\mu} \phi  \right)^* \left( \phi^\dagger D^{\mu} \phi \right).
\end{split}
\end{equation}
The interactions with the physical (neutral) Higgs boson $h$ in the SMEFT in the unitary gauge are obtained setting $\phi=\frac{1}{\sqrt{2}} \left(
0,v_\mathrm{T}+h[1+\adcoeff{\mathrm{kin}}{}] \right)^T$ in Eq.~\eqref{eq:Operators}. 
Higher dimensional operators produce $\order{v^2/\Lambda^2}$ corrections to the vacuum expectation value and give rise to a non-canonically normalised Higgs kinetic term. The latter can be put back in the canonical normalisation by means of a redefinition of the Higgs field, yielding a universal shifts of all the single Higgs couplings
(see Ref.~\cite{Jenkins:2017jig} for more details):
\begin{equation}
\begin{split}
v_\mathrm{T} &= \left(1 + \frac{3 \coeff{\phi}{} v^2}{8 \lambda} \right) v,\\
\adcoeff{\mathrm{kin}}{} &= \left( \coeff{\phi \Box}{} - \frac{1}{4} \coeff{\phi D}{} \right) v^2.
\end{split}
\end{equation}
The terms in Eq.~\eqref{eq:Operators} involving bosons and zero or two quarks are collected in
\begin{equation}
\begin{split}
\lag_{\mathrm{B}}  &=  
 \coeff{G}{} f^{ABC} G_{\mu} ^{A,\nu} G_{\nu} ^{B,\rho} G_{\rho} ^{C,\mu} + \coeff{\phi G}{} \left( \phi^\dagger \phi \right) G_{\mu \nu}^A G^{A,\mu\nu}  \\
& + \coeff{t \phi}{} \left[ \left( \phi^\dagger \phi \right) \bar{Q}_L \Tilde{\phi} t_R  + \hc\right] \\ 
&+ \coeff{tG}{}  \left[\left( \bar{Q}_L \sigma^{\mu \nu} T^A t_R \right)  \Tilde{\phi} G_{\mu \nu} ^A + \hc \right].
\end{split}
\label{eq:L2t}
\end{equation}
While we follow Ref.~\cite{dim6smeft} in the notation of the Wilson coefficients, it should be noticed that we are performing a slight change of notation: since we are focusing on the third generation, we fix $\coeff{tG}{} = \coeff{uG}{33}$ (and in analogy for the other Wilson coefficients).

In the following, we assume the Wilson coefficients to be real, neglecting the possibility of CP violation. This choice is justified by the severe bounds on CP violating SMEFT operators from electric dipole moment, see Refs.~\cite{Chien:2015xha,Cirigliano:2016njn}.\footnote{While some combinations of CP violation in different Higgs couplings can survive the bounds, LHC searches (see Refs.~\cite{Bahl:2020wee, Martini:2021uey, Bahl:2021dnc}) also in $t\bar{t}h$ can provide complementary probes. We do not consider this possibility though in the following.}

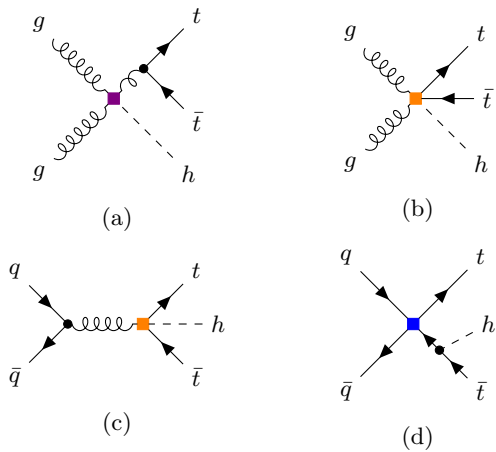
\begin{figure}[]
    \centering
    \begin{subfigure}[t]{0.45\linewidth}
        \centering
        	\begin{tikzpicture}[baseline=(a)]
				\begin{feynman}[small]
					\vertex (a) [violet,square dot,scale=\sizesqdot] {};
					\vertex  (gi2) [below left=40 pt of a]{\( g \)};
					\vertex  (gi1) [above left=40 pt of a]{\( g \)};
					
					\vertex (h) [below right = 40 ptof a] {\( h \)};
					\vertex (g) [dot, scale = \sizedot,above right = 16 pt of a] {};
					\vertex  (t1) [above right = of g]{\( t \)};
					\vertex  (t2) [below right = of g]{\( \bar{t} \)};
					\diagram* {
						(gi1)  -- [gluon] (a),
						(gi2)  -- [gluon] (a),
						(g)  -- [gluon] (a),
						(h) -- [scalar] (a),
						(t2) --[fermion] (g) --[fermion] (t1)
					};
				\end{feynman}
			\end{tikzpicture}
        \caption{}\label{fig:ggtthSMEFT1}
    \end{subfigure}
      \begin{subfigure}[t]{0.45\linewidth}
        \centering
        	\begin{tikzpicture}[baseline=(a)]
				\begin{feynman}[small]
					\vertex (a) [orange,square dot,scale =\sizesqdot] {};
					\vertex  (gi2) [below left=35 pt of a]{\( g \)};
					\vertex  (gi1) [above left= 35 pt  of a]{\( g \)};
					\vertex (h) [below right = 35 pt  of a] {\( h \)};
					\vertex  (t1) [above right = 35 pt  of a]{\( t \)};
					\vertex  (t2) [right = 27 pt   of a]{\( \bar{t} \)};
					\diagram* {
						(gi1)  -- [gluon] (a),
						(gi2)  -- [gluon] (a),
						(t2) --[fermion] (a) --[fermion] (t1),
						(h) --[scalar] (a)
					};
				\end{feynman}
			\end{tikzpicture}
        \caption{}\label{fig:ggtthSMEFT2}
    \end{subfigure}
    \begin{subfigure}[t]{0.45\linewidth}
        \centering
        \begin{tikzpicture}[baseline=(h)]
            \begin{feynman}[small]
                \vertex (q1) {\(  q \)};
                \vertex (qqg1) [dot, scale=\sizedot, below right=of q1] {};
                 \vertex (q2) [below left =  of qqg1] {\( \bar{q} \)};
                \vertex (qqg2) [orange, square dot, scale=\sizesqdot,right = of qqg1] {};
                \vertex (h) [right = of qqg2] {\( h \)};                
                \vertex  (t1) [above right= of qqg2]{\(  t \)};
                \vertex  (t2) [below right= of qqg2]{\(  \bar{t} \)};
                \diagram* {
                    (q1) --[fermion] (qqg1) -- [fermion] (q2),
                    (qqg1) --[gluon] (qqg2),
                    (t2) -- [fermion] (qqg2)  -- [fermion] (t1),
                    (h) -- [scalar] (qqg2)
                };
            \end{feynman}
        \end{tikzpicture}
        \caption{}\label{fig:qqtthSMEFT1}
    \end{subfigure}
      \begin{subfigure}[t]{0.45\linewidth}
        \centering
        \begin{tikzpicture}[baseline=(h)]
            \begin{feynman}[small]
                \vertex (q1) {\(  q \)};
                \vertex (qqg1) [blue,square dot, scale=\sizesqdot, below right=36 pt of q1] {};
                 \vertex (q2) [below left =36 pt  of qqg1] {\( \bar{q} \)};
                \vertex (h) [right= of qqg1] {\( h \)};                
                \vertex (tth) [dot, scale=\sizedot,below right = 14 pt of qqg1] {};
                \vertex  (t1) [above right=36 pt of qqg1]{\(  t \)};
                \vertex  (t2) [below right=36 pt of qqg1]{\(  \bar{t} \)};
                \diagram* {
                    (q1) --[fermion] (qqg1) -- [fermion] (q2),
                    (t2) --[fermion] (tth) -- [fermion] (qqg1)  -- [fermion] (t1),
                    (h) -- [scalar] (tth)
                };
            \end{feynman}
        \end{tikzpicture}
        \caption{}\label{fig:qqtthSMEFT2}
    \end{subfigure}
\caption{Tree-level diagrams for $pp\to t\bar{t}h$ in the SMEFT. The purple square denotes an insertion of $\op{\phi G}{}$, the orange square denotes an insertion of $\op{tG}{}$ and the blue square denotes an insertion of $\lag_{\mathrm{4F}}$.} \label{fig:pptthSMEFT}
    \end{figure}

Finally, $\lag_{\mathrm{4F}}$ contains all the four-quark operators in the Warsaw basis having a non-vanishing matrix element for $\bar{q}q\bar{t}{t}$. Clearly these terms affect only the $\bar{q}q \to t\bar{t}h$ channel (at tree level).\par
\begin{equation} \label{eq:L4F}
\begin{split}
 \lag_{\mathrm{4F}} &= \coeff{uu}{prst} \left( \bar{u}_R^p\gamma_\mu u_R^r \right)\left( \bar{u}_R^s\gamma^\mu u_R^t \right) \\
 &+\coeff{qq(1)}{prst} \left( \bar{q}_L^p\gamma_\mu q_L^r \right)\left( \bar{q}_L^s\gamma^\mu q_L^t \right) \\ &+\coeff{qq(3)}{prst} \left( \bar{q}_L^p \tau^I \gamma_\mu q_L^r \right)\left( \bar{q}_L^  s  \tau^I \gamma^\mu q_L^t \right) \\
 &+\coeff{qu(1)}{prst} \left( \bar{q}_L^p\gamma_\mu q_L^r \right)\left( \bar{u}_R^s\gamma^\mu u_R^t \right) 
 \\ &+ \coeff{qu(8)}{prst} \left( \bar{q}_L^p T^A \gamma_\mu q_L^r \right)\left( \bar{u}_R^s T^A\gamma^\mu u_R^t \right) \\
 &+\coeff{qd(1)}{prst} \left( \bar{q}_L^p\gamma_\mu q_L^r \right)\left( \bar{d}_R^s\gamma^\mu d_R^t \right) 
\\ &+\coeff{qd(8)}{prst} \left( \bar{q}_L^p T^A \gamma_\mu q_L^r \right)\left( \bar{d}_R^s T^A\gamma^\mu d_R^t \right) \\
 &+ \coeff{quqd(1)}{prst} \left( \bar{q}_L^{j,p} u_R^r \right) \varepsilon_{jk} \left( \bar{q}_L^{k,s} u_R^t \right) + \hc \\
 &+ \coeff{quqd(8)}{prst} \left( \bar{q}_L^{j,p} T^A u_R^r \right) \varepsilon_{jk} \left( \bar{q}_L^{k,s} T^A u_R^t \right) + \hc .
\end{split}
\end{equation}
In the above expression, $p,r,s,t$ are the generation indices, $j,k$ are the $\group{SU(2)}_\mathrm{L}$ indices in the fundamental representation and $\tau^I$ are the Pauli matrices. It holds $ Q_L \equiv q^3_L$, $t_R \equiv u_R^3$. Obviously, we are interested in the combinations giving a non-zero matrix element between light quarks and top quarks (e.g.~$\op{uu}{1133}$).
A sample of the tree-level diagrams arising due to higher dimensional operators is given in Fig.~\ref{fig:pptthSMEFT}.

In this work, we do not consider contributions where the Higgs boson couples to light quarks and gluons because such operators are not expected to show large running effects, being suppressed by light Yukawa couplings. 

In Eqs.~\eqref{eq:Lphi},\eqref{eq:L2t},\eqref{eq:L4F} we include only operators which enter at tree-level in $pp \to t\bar{t}h$. This choice does not take into account the fact that some operators (such as the chromomagnetic operator) cannot be generated at tree-level, assuming that the UV completion of the SM is weakly coupled and renormalisable (see Ref.~\cite{Arzt:1994gp,Buchalla:2022vjp}). In other words, Eqs.~\eqref{eq:Lphi},\eqref{eq:L2t},\eqref{eq:L4F} follow an expansion based only on the canonical dimension of the operators.

Four-top operators enter $pp\to t\bar{t}h$ at one-loop level. Schematically, their (bare) contribution goes as $\mtrx{}\sim \coeff{\mathrm{4t}}{} \times  1/(16 \pi^2) \times \left( A+B\left(1/\epsilon + \log \mu_{\mathrm{R}}^2/\Lambda^2 \right) \right)$. Since the poles determine the anomalous dimension of the theory, including running effects in operators which enter at tree-level in the process means resumming the logarithmic terms. 
The finite terms represented by $A$ can be phenomenologically relevant, as described in Ref.~\cite{Alasfar:2022zyr}. However, we do not include such terms in our analysis since they do not depend on the choice of the renormalisation scale $ \mu_{\mathrm{R}}$ and do not contribute to the difference of the result obtained with a fixed or dynamical renormalisation scale. However, for a precise phenomenological analysis, they should be included. In the same spirit, we do not include the SM NLO QCD and electroweak corrections but note that they moderately increase the cross section, see Refs.~\cite{Dawson:2003zu, Beenakker:2002nc, Dawson:2003zu, Beenakker:2002nc, Zhang:2014gcy, Frixione:2014qaa}.

As it has been pointed out recently in Ref.~\cite{DiNoi:2023ygk}, there is a non-trivial interplay between four-fermion operators and other SMEFT operators. Depending on of the continuation scheme chosen for the $\gamma_5$ in $D\ne 4$ space-time dimensions a four-fermion operator generated at tree-level, e.g. $\op{qu(1)}{3333}$, automatically introduces a one-loop coefficient to e.g.~$\op{tG}{}$. We will for the moment refrain from considering this interplay but come back to this in Scenario 2 analysed in Sec.~\ref{sec:results}. We note for the moment that the coefficients of the operators considered in this work can be considered independent when using the \textit{Breitenlohner-Maison-’t Hooft-Veltman} (BMHV) scheme for $\gamma_5$ (see Refs.~\cite{THOOFT1972189, Breitenlohner:1977hr}). 

We compute matrix elements at $\order{\frac{1}{\highscale^2}}$ and we consider cross sections at $\order{\frac{1}{\highscale^4}}$.\footnote{
We choose to include $\order{\frac{1}{\highscale^4}}$ terms in the matrix element squared (stemming from dimension-six operators squared)
as we require a positive cross section in the whole phase space. We note though that, if the cross section beecomes negative when considering $\order{\frac{1}{\Lambda^2}}$ terms only, the EFT validity might be questionable.
}
In other words, we consider only diagrams with a single insertion of a dimension-six operators but we do not include interference terms between the SM and dimension-eight operators. We generate the diagrams with {\tt qgraf-3.6.5} \cite{Nogueira:1991ex} and perform the algebra with {\tt FeynCalc} \cite{Mertig:1990an, Shtabovenko:2016sxi,Shtabovenko:2020gxv}.

While the squaring of the matrix element for the quark-induced channels is straightforward, the sum over external polarisation states of the gluons 
\begin{equation} \label{eq:pol1}
\sum_{\mathrm{Pol}} \epsilon_\alpha (p)\epsilon_\beta(p)^* = - g_{\alpha\,\beta} +\frac{n_\alpha p_\beta}{(n \cdot p)} + \frac{p_\alpha n_\beta}{(n \cdot p)} -  \frac{p_\alpha p_\beta}{(n \cdot p)^2}
\end{equation}
is computationally expensive (being $n$ a generic unit vector which must drop from the final result). We used a shortcut, based on the optical theorem (see Ref.~\cite{externalgluons} for the details). It is possible to use the simple expression $\sum_{\mathrm{Pol}} \epsilon_{\alpha} (p)\epsilon_{\beta}(p)^* = - g_{\alpha\,\beta}$ if we subtract incoherently the squared matrix element where the gluons are replaced by the ghosts ($\eta$).
Let $\mtrx{gg \to t\bar{t} h} =\epsilon_{\mu_1} (q_1 ) \epsilon_{\mu_2} (q_2) A^{\mu_1 \mu_2}$: the unpolarized squared matrix element is given by
\begin{equation}\label{eq:pol2}
\begin{split}
\sum_{\mathrm{Pol}}|\mtrx{gg \to t\bar{t} h}|^2 &=g_{\mu_1 \nu_1} g_{\mu_2 \nu_2} A^{\mu_1 \mu_2}  \left(A^{\nu_1 \nu_2} \right)^* \\ 
&- 2 |\mtrx{\bar{\eta} \eta \to t\bar{t} h}|^2.
\end{split}
\end{equation} 
We explicitely verified the correctness of this approach using the SM amplitude for $gg \to t\bar{t}h$. 

For the numerical evaluation of the differential cross section we rely on our own code written in \texttt{Fortran}, \texttt{C++} and \texttt{python}.
The cross section of the $i$-th event is given by 
\begin{equation}
\sigma_i = \sum_{X} \frac{\mathfrak{S}_X}{2 \hat{s}}  \,  \, w_X(\mu_\mathrm{F},x_1,x_2) \left( \sum_{\mathrm{Pol}}|\mtrx{X \to t\bar{t} h}|^2 \right)_i \, \mathrm{PS}_i,
\end{equation}
with $\hat{s} = x_1 x_2 E_{\mathrm{Collider}}^2\ge (2 m_t+m_h)^2$, where $m_t$ is the top quark mass and $m_h$ the Higgs mass. The phase space $\mathrm{PS}_i$ is generated by \texttt{rambo} \cite{rambo} and $X=gg,\bar{u}{u},\bar{d}{d}...$ denote the different partonic channels. The average over the initial states yelds $\mathfrak{S}_{gg}=\frac{1}{256}$ and $\mathfrak{S}_{\bar{q}q}=\frac{1}{36}$ and we set the collider energy to $E_{\mathrm{Collider}} = 13.6 \, \mathrm{TeV}$ throughout the paper. 
The weight of the parton distribution function (PDF) is given by
\begin{equation} \label{eq:PDFs}
    \begin{split}
     w_{gg}(\mu_\mathrm{F},x^i_1,x^i_2) &= f_{g}(\mu_\mathrm{F},x_1^i) f_{g}(\mu_\mathrm{F},x_2^i), \\
     w_{\bar{q} q}(\mu_\mathrm{F},x^i_1,x^i_2) &= f_{\bar{q}}(\mu_\mathrm{F},x_1^i) f_{q}(\mu_\mathrm{F},x_2^i)\\
     &+f_{\bar{q}}(\mu_\mathrm{F},x_2^i) f_{q}(\mu_\mathrm{F},x_1^i),
    \end{split}
    \end{equation}
being $\mu_\mathrm{F}$ the factorisation scale set to 
$\mu_\mathrm{F}=H_{T}/2$, with $H_{T} \equiv p_{T,t}+p_{T,\bar{t}}+p_{T,h}$. 
    
We access the PDFs via \texttt{LHAPDF-6.5.3} \cite{lhapdf}, employing the PDF set \texttt{NNPDF40\_lo\_as\_01180} \cite{NNPDF:2021njg}.

The total cross section is given by
\begin{equation}
\sigma(pp \to t\bar{t}h) = \frac{1}{N_{\mathrm{E}}} \sum_{i=1}^{N_{\mathrm{E}}} \sigma^i,
\end{equation}
where $N_{\mathrm{E}}$ denotes the number of events. 
The errors are computed as the square root of the bin sum of the $\sigma^i$.

Using a dynamical scale implies that the Wilson coefficients of the SMEFT operators should be evaluated at this scale. Given the definition of the set of non-vanishing SMEFT operators at the scale $\mu = \highscale$, $\{ \coeff{i}{} (\highscale)\}$, we have to solve the RGEs from $\highscale$ to $\mu_{\mathrm{R}}$ in order to obtain $\{ \coeff{i}{} (\mu_{\mathrm{R}})\}$ and use them to compute the matrix element squared. We stress that this must be done for every event if one chooses a dynamical scale: high time efficiency is crucial in this context. 
To reduce the computational time is it possible to rely on some approximations, for instance like solving the RGEs analytically only for the dominant contribution (typically $\alpha_s$, see Ref.~\cite{Maltoni:2016yxb,Aoude:2022aro,Grazzini:2018eyk, Battaglia:2021nys}) or employing the first leading-log approximation,
\begin{equation}
\label{eq:approximate}
\coeff{i}{}(\mu_{\textrm{R}}) = \coeff{i}{}(\highscale) +   
\adm_{ij}(\highscale) \coeff{j}{}(\highscale)  \frac{\log{\left( \mu_{\textrm{R}} / \highscale\right)}}{16 \pi^2},
	\end{equation}
with $\mu \frac{ d \coeff{i}{}}{d \mu} = \frac{1}{16 \pi^2}\adm_{ij}\coeff{j}{}$. The first leading-log approximation is expected to provide an accurate result only if $\adm_{ij}(\highscale) \coeff{j}{}(\highscale) {\log{\left( \mu_{\textrm{R}} / \highscale\right)}}/{(16 \pi^2)}$ is small enough to remain inside the perturbative regime. 
We will come back on this topic in Scenario 3 in Sec.~\ref{sec:results}.

We solve the RGEs numerically using \texttt{RGESolver} version \texttt{v1.0.2}, presented in Ref.~\cite{rgesolver}.\footnote{Other publicly available codes performing the running in the SMEFT are \texttt{DSixTools2.0} \cite{Fuentes-Martin:2020zaz} and \texttt{wilson} \cite{Aebischer:2018bkb}.} The numerical approach allows us to consider not only $\order{\alpha_s}$ effects but also $\order{\yuk{t}}$ effects and address their importance.

Differential cross sections are obtained by respective binning of the events. The initial conditions for the SM parameters are obtained evolving the pure SM RGEs up to $\Lambda$ using the default input provided by \texttt{RGESolver} (meaning Table~4 of Ref.~\cite{rgesolver}).
In the following, we distinguish between the three set-ups. In all these cases, we evolve from $\mu=\Lambda$ to $\mu=\mu_{\mathrm{R}}$. 
\begin{itemize}
    \item SMEFT (Numeric RGE): a dynamical scale $\mu_\mathrm{R}=\mu_\mathrm{F}=H_T/2$ is used (relying on numeric solution of the one-loop RGEs).
    \item SM: same as before but setting at $\mu = \Lambda$ all the SMEFT coefficients to 0. This choice represents the benchmark with respect to which we compare the SMEFT effects, meaning that we use for the SM input parameters and their running exactly the same input as for the SMEFT case.
    \item SMEFT (Fixed Scale): a fixed scale $\mu_{\mathrm{R}}=m_t$ is used for all the parameters of the theory. To isolate the contribution coming from pure SMEFT running, the strong coupling $\alpha_s$ is still evaluated at the dynamical scale $\mu_{\mathrm{F}}=H_T/2$. Further discussion regarding this aspect can be found in the next section for the operator $\op{Qt}{(1)}$.
\end{itemize}
The masses are renormalised on-shell: we use $m_t = 173\,\mathrm{GeV}$, $m_h=125 \, \mathrm{GeV}$. The effective coupling of Higgs to top quarks is given by:
\begin{equation} \label{eq:ghtt}
g_{h t\bar{t}}(\mu_{\mathrm{R}}) = \frac{m_t}{v_{\mathrm{T}}} (1+\adcoeff{\mathrm{kin}}{}(\mu_{\mathrm{R}}))  -\frac{3 v^2}{2 \sqrt{2}}  \coeff{t \phi}{}(\mu_{\mathrm{R}}) .
\end{equation}
The vacuum expectation value is not directly measured. It is derived from $G_F$, which is measured in muon decay. When SMEFT operators are considered, this input value is modified by the operators $\op{Hl(3)}{11,22}$ and $\op{ll}{1221}$ (see Ref.~\cite{Brivio:2020onw}). We neglect this effect since such operators do not play any role in our analysis, see Refs.~\cite{Corbett:2021cil,Martin:2023fad, Biekotter:2023xle, Biekotter:2023xxx} for detailed analyses.

\section{Results} \label{sec:results}
We show the dependence of the differential cross section on the running of the Wilson coefficients in three different scenarios. All the plots displayed in this section have been obtained simulating $10^5$ events. The whole computation (event generation, running and matrix element evaluation) takes $\order{20 \, \mathrm{min}}$ on a normal laptop.

We show in many plots the percentual difference between the fixed and dynamical scale in each bin, namely
\begin{equation}
\Delta \equiv \frac{
\left( \frac{d \sigma }{d p_{T,h}} \right)_{\mathrm{Dyn}}-\left( \frac{d \sigma }{d p_{T,h}} \right)_{\mathrm{Fixed}}
}{\left( \frac{d \sigma }{d p_{T,h}} \right)_{\mathrm{Fixed}}
}. \label{eq:percent}
\end{equation}

\paragraph{Scenario 1}: We turn on one Wilson coefficient at the time, using $\highscale=2\,\mathrm{TeV}$.
The idea of this scenario is to demonstrate the importance of $\yuk{t}$ running with respect to pure $\alpha_s$ running. 
To do so, we compare two cases: we turn on individually the four-top operators $\op{Qt}{(1,8)} \equiv \op{qu(1,8)}{3333}$, choosing as initial conditions (i) $\coeff{Qt}{(1)} = (4/3) \times 20 \, / \mathrm{TeV}^2$ and (ii) $\coeff{Qt}{(8)} = 20 \, / \mathrm{TeV}^2$. This choice, consistent with the individual bounds at $\order{1/\Lambda^2}$ presented in Ref.~\cite{Ethier:2021bye}, is motivated by the fact that these operators enter the $\beta$-functions of $\coeff{t \phi}{}$ with pure Yukawa-induced terms in the combination $\left(\coeff{Qt}{(1)}+ C_F \coeff{Qt}{(8)} \right)$, being $C_F=4/3$ for $\group{SU(3)}_{\mathrm{C}}$ (see Refs.~\cite{rge1,rge2}).
This concretely means that both choices (i) and (ii) generate the same contribution to the running of $\coeff{t \phi}{}$ at leading-logarithm, hence the same Yukawa-induced running in both cases. However, $\op{Qt}{(8)}$ contributes to the $\alpha_s$ running of other operators entering at tree-level in $pp \to t\bar{t}h$ via, for example, penguin diagrams (e.g.~it generates $\coeff{uu}{1133} (\mu_{\mathrm{R}})\ne 0$). This is not the case for $\op{Qt}{(1)}$. For this reason, a comparison between the two operators provides a good indicator of the importance of purely Yukawa-induced RGE running effects with respect to the ones proportional to the strong coupling for this process.

\begin{figure*}[t]
  \centering
  \begin{minipage}[b]{0.48\textwidth}
    \centering
    \includegraphics[width=\linewidth]{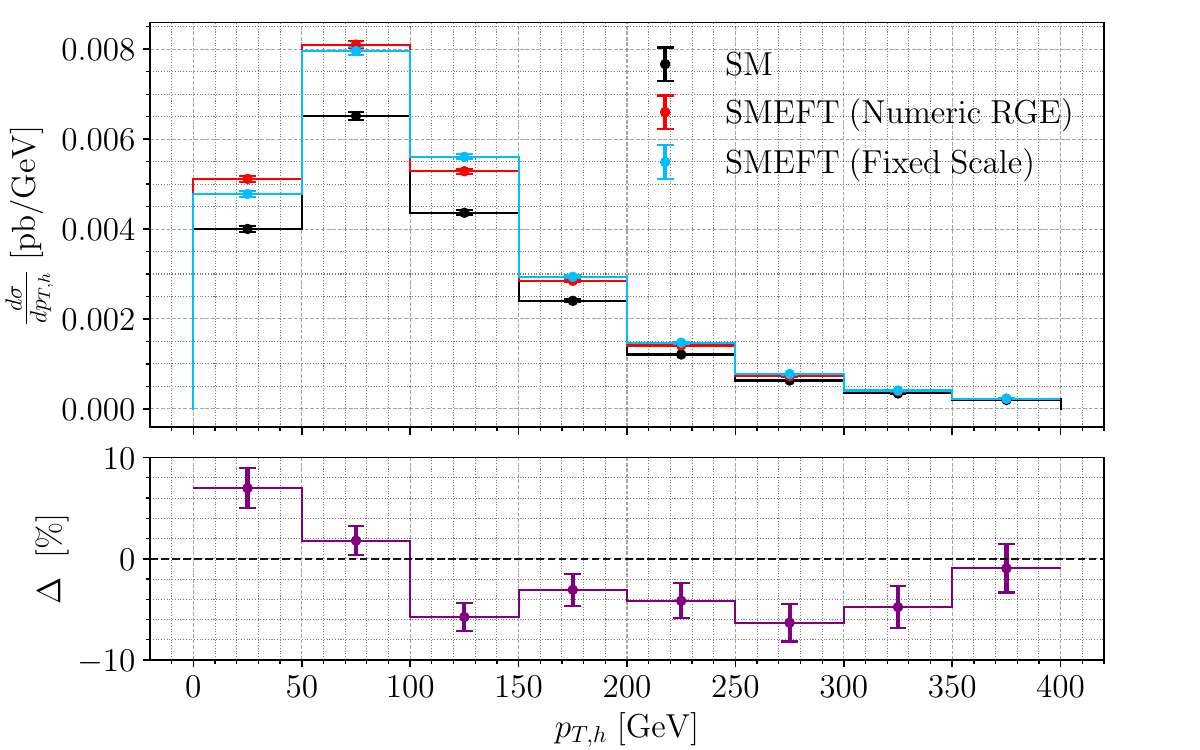}
    \caption{Higgs transverse momentum distribution (upper panel) and percentual difference for each bin between the dynamical scale and the fixed scale (lower panel) with $\coeff{Qt}{(1)}(\Lambda) = \frac{4}{3} \times 20 \, / \mathrm{TeV}^2$.}
    \label{fig:dsigmaQt1}
    \includegraphics[width=\linewidth]{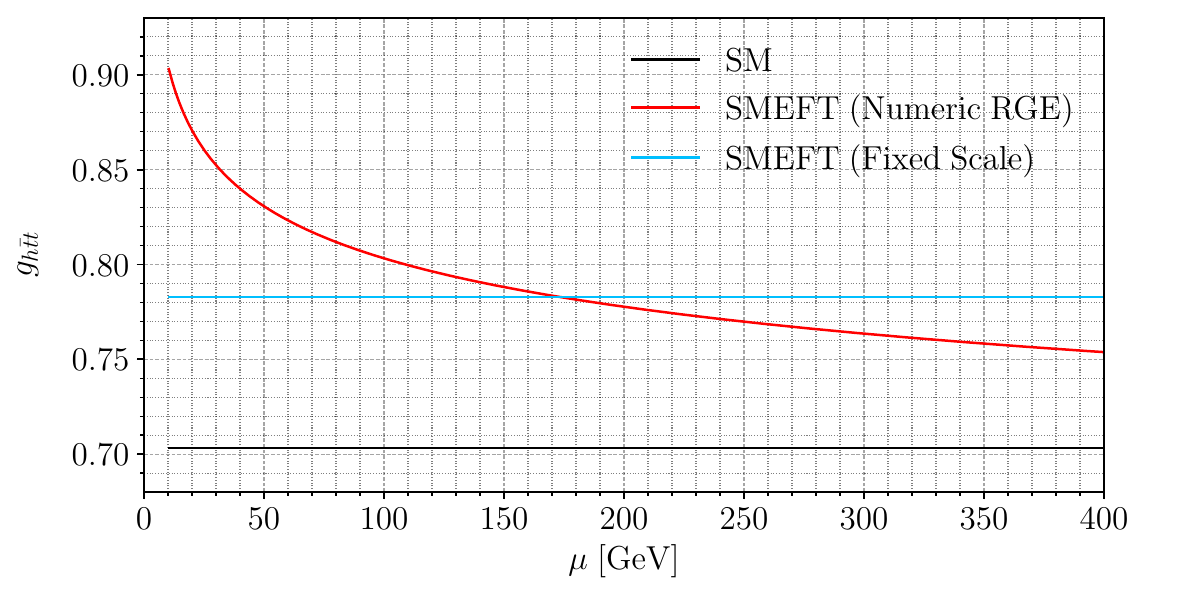}
\caption{Running of $g_{h t\bar{t}}$ with $\coeff{Qt}{(1)}(\Lambda) = \frac{4}{3} \times 20 \, / \mathrm{TeV}^2$.}
\label{fig:ghttRunningQt1}
  \end{minipage}\hfill
   \begin{minipage}[b]{0.48\textwidth}
    \centering
    \includegraphics[width=\linewidth]{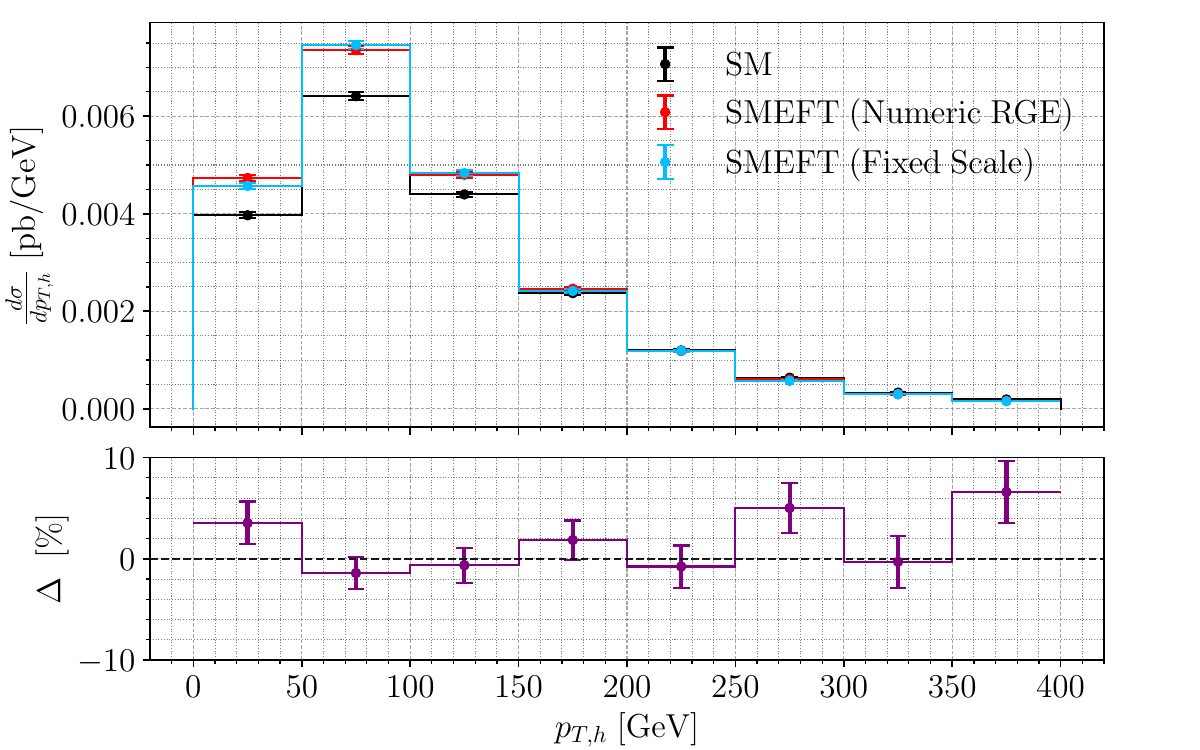}
    \caption{Higgs transverse momentum distribution (upper panel) and percentual difference for each bin between the dynamical scale and the fixed scale (lower panel) with $\coeff{Qt}{(8)}(\Lambda) = 20 \, / \mathrm{TeV}^2$.}
    \label{fig:dsigmaQt8}
      \includegraphics[width=\linewidth]{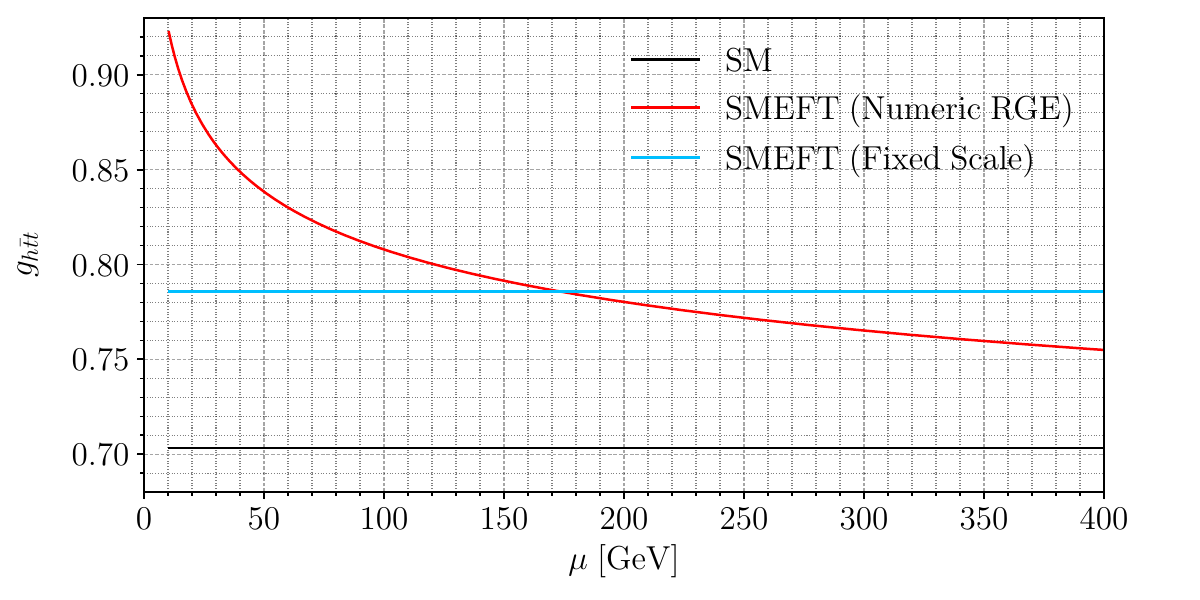}
\caption{Running of $g_{h t\bar{t}}$ with $\coeff{Qt}{(8)}(\Lambda) = 20 \, / \mathrm{TeV}^2$.}
\label{fig:ghttRunningQt8}
  \end{minipage}
\end{figure*}

We report the differential distribution with respect to the Higgs transverse momentum $p_{T,h}$ for $\coeff{Qt}{(1)}$ in Fig.~\ref{fig:dsigmaQt1} and for $\coeff{Qt}{(8)}$ in Fig.~\ref{fig:dsigmaQt8}. 

We observe in both cases differences in the distribution obtained with the dynamical scale with respect to the one obtained with the fixed scale below $10 \%$. The two choices of $\mu_{\mathrm{R}}$ agree rather well in the $[50,100]\, \mathrm{GeV}$ bin. In both cases, we see a small enhancement at low $p_{T,h}$, which can be qualitatively understood looking at the energy-scale dependence of $g_{ht\bar{t}}$, displayed in Fig.~\ref{fig:ghttRunningQt1} (Fig.~\ref{fig:ghttRunningQt8}) for $\coeff{Qt}{(1)}$ ($\coeff{Qt}{(8)}$). 
The small difference between Fig.~\ref{fig:ghttRunningQt1} and Fig.~\ref{fig:ghttRunningQt8} can be attributed to the non leading-log contributions in the running of $\coeff{t \phi}{}$. 

Since $\op{Qt}{(1)}$ does not mix with operators contributing at tree-level to $pp \to t\bar{t}h$ via QCD interactions, the fact that we find a (small) deviation between fixed and dynamical scale means that purely top Yukawa-induced running has an effect on the distribution. For this reason, it should be included in presence of large Wilson coefficients. Indeed, we find in this case that both the operators with Wilson coefficients $\op{Qt}{(1)}$  and $\op{Qt}{(8)}$ show deviations between fixed and dynamical scale of the same order, which lets us conclude that the effects stemming from top Yukawa-induced running can be as important as the ones propotional to $\alpha_s$.

\begin{figure}
    \centering
    \includegraphics[width=\linewidth]{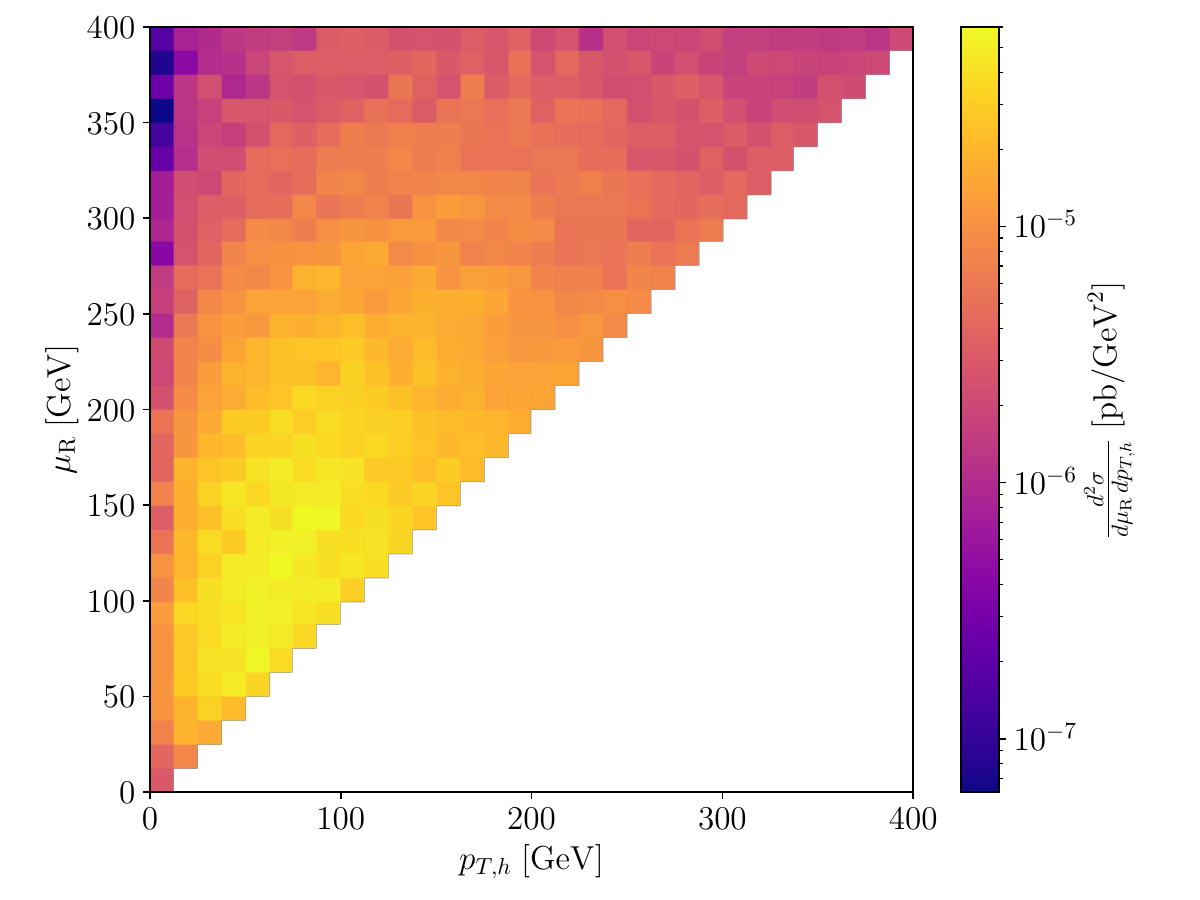}
    \caption{Correlation between $p_{T,h}$ and $\mu_{\mathrm{R}}$ for $\coeff{Qt}{(1)}(\Lambda) = (4/3) \times 20 \, / \mathrm{TeV}^2$.}
    \label{fig:d2sigmaQt1}
\end{figure}

It is also interesting to look at the correlation between $p_{T,h}$ (which is a physical observable) and $\mu_{\mathrm{R}}$ (which is an arbitrary energy scale). We quantify this correlation looking at $\frac{d^2 \sigma}{d \mu_{\mathrm{R}} \, d p_{T,h}}$, reporting the result for $\coeff{Qt}{(1)}$ in Fig.~\ref{fig:d2sigmaQt1}. We expect roughly the same result for the bin $p_{T,h} \in [50,100]\times \mathrm{GeV}$, since most of the events falling in such interval produce $\mu_{\mathrm{R}} \sim m_t$. This is precisely what can be inferred from Fig.~\ref{fig:d2sigmaQt1}. We also note that the largest part of $\frac{d^2 \sigma}{d \mu_{\mathrm{R}} \, d p_{T,h}}$ comes from values of $p_{T,h} \sim \mu_{\mathrm{R}}$. 

\begin{figure}[t]
\centering
    \includegraphics[width=\linewidth]{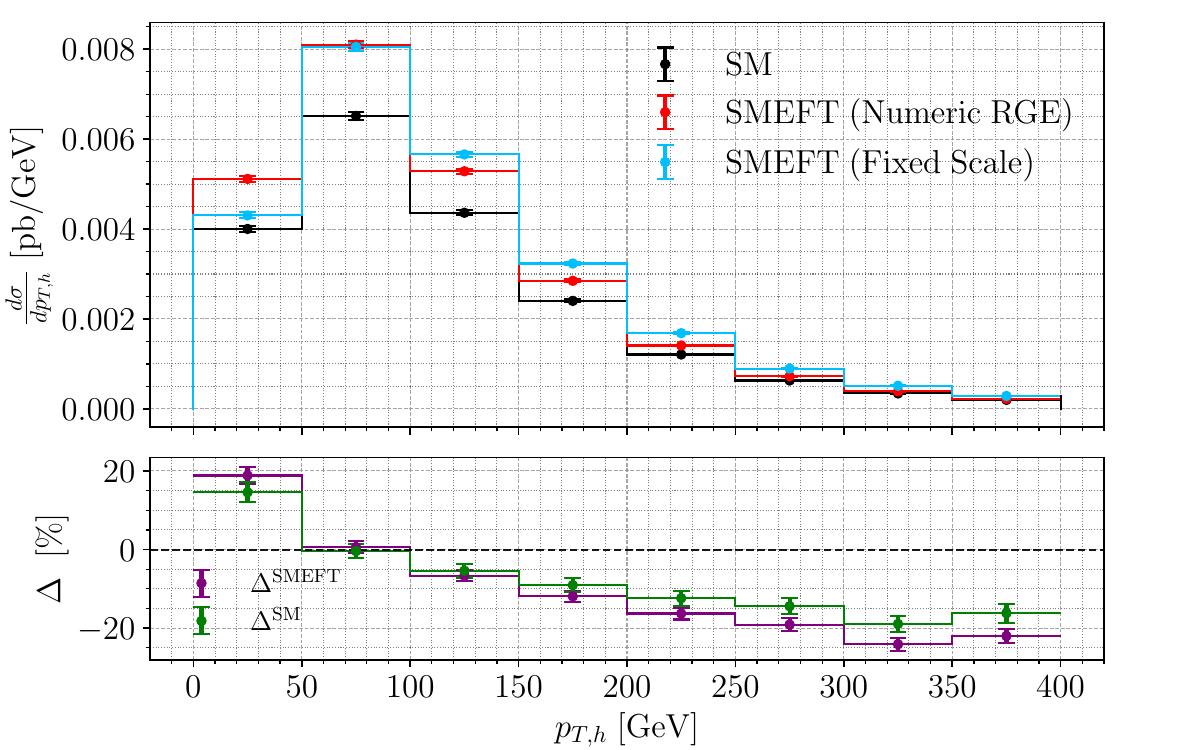}
    \caption{Same than Fig.~\ref{fig:dsigmaQt1} but with $\alpha_s$ evaluated at $\mu_{\mathrm{R}}=m_t$ as the SMEFT coefficients for the fixed-scale scenario, see text for the details.}
    \label{fig:dsigmaQt1gsRun}
\end{figure}

Lastly, we comment about the choice of using a different renormalisation scale for the strong coupling ($\mu_{\mathrm{R}}=\mu_{\mathrm{F}}=H_T/2$) and for the SMEFT coefficients ($\mu_{\mathrm{R}}=m_t$) in the fixed-scale case. This choice allows to highlight the importance of the running of the SMEFT coefficients, separating it from pure SM running of $\alpha_s$. For the sake of clarity, we also report in Fig.~\ref{fig:dsigmaQt1gsRun} the results for $\coeff{Qt}{(1)}$ in which the renormalisation scale in the fixed-scale scenario is set to $\mu_{\mathrm{R}}=m_t$ also for the strong coupling. With this choice, we can see a larger difference between the fixed scale and the dynamical scale distribution. However, part of this difference is due to pure SM strong running, as can be inferred from the lower panel of the same figure. 

\paragraph{Scenario 2}: We use a UV model inspired setting of the Wilson coefficients at a high scale. We add to the SM a new scalar $\Phi \sim (8,2)_{\frac{1}{2}}$ with a mass $M_{\Phi} \gg v$:
\begin{equation}
\begin{split}
\mathcal{L}_\Phi &= (D_\mu \Phi)^{\dagger}  D^\mu \Phi - M_\Phi^2 \Phi^\dagger \Phi \\ &- Y_{\Phi}  \left( \Phi^{A,\dagger} \varepsilon \bar{Q}_L ^T  T^A t_R + \hc \right).
\end{split}
\end{equation}
The tree-level matching, performed at $\mu = \Lambda$, gives rise to (see Ref.~\cite{deBlas:2017xtg}):
\begin{equation}\label{eq:4tmatchingOctet}
{\coeff{Qt}{(1)}} = -\frac{2}{9} \frac{Y_{\Phi}^2}{ M_\Phi^2}, \quad {\coeff{Qt}{(8)}} = \frac{1}{6} \frac{Y_{\Phi}^2}{ M_\Phi^2}.
\end{equation}
At one-loop level the chromomagnetic operator can be generated.\footnote{Together with it, one-loop shifts to Fierz identities induce also other operators, see Ref.~\cite{Fuentes-Martin:2022jrf}. However, neglecting all the Yukawa couplings other than the top quark one, the only operator relevant for our process generated at loop-level is the chromomagnetic operator.} The value of its coefficient depends on the continuation scheme chosen for $\gamma_5$:
\begin{equation}
\mathcal{C}_{tG}^{\mathrm{NDR}}= \frac{1}{16 \pi^2} \frac{\yuk{\Phi}^2}{ M_\Phi^2}\frac{g_s \yuk{t}}{4}, \quad \mathcal{C}_{tG}^{\mathrm{BMHV}}= 0.     \end{equation}
The index NDR stands for na\"ive dimensional regularisation. We refer to its exact definition to Ref.~\cite{DiNoi:2023ygk} which also discusses exhaustively the scheme-independence of the final result. Relevant for the following discussion is the fact that in the NDR scheme the four-top operators contribute at two-loop order to the running of $\coeff{\phi G}{}$. Its $\beta$-function is thus given by 
\begin{equation}
16 \pi^2 \mu \frac{d \coeff{\phi G}{}}{d \mu} = -4 g_s \yuk{t}  \left( \coeff{tG}{} + \frac{ g_s\yuk{t}}{16 \pi^2}\left(\coeff{Qt}{(1)} - \frac{1}{6} \coeff{Qt}{(8)} \right) \right).
\label{eq:RGECphiG}
\end{equation}
We note that the two terms are indeed of the same order, since $\op{tG}{}$ is generated at one-loop level in the matching. If one would adopt a loop counting of the operators this would be directly manifest within the RGEs.

\begin{figure}[]
  \centering
  \begin{subfigure}[b]{\linewidth}
    \includegraphics[width=\linewidth]{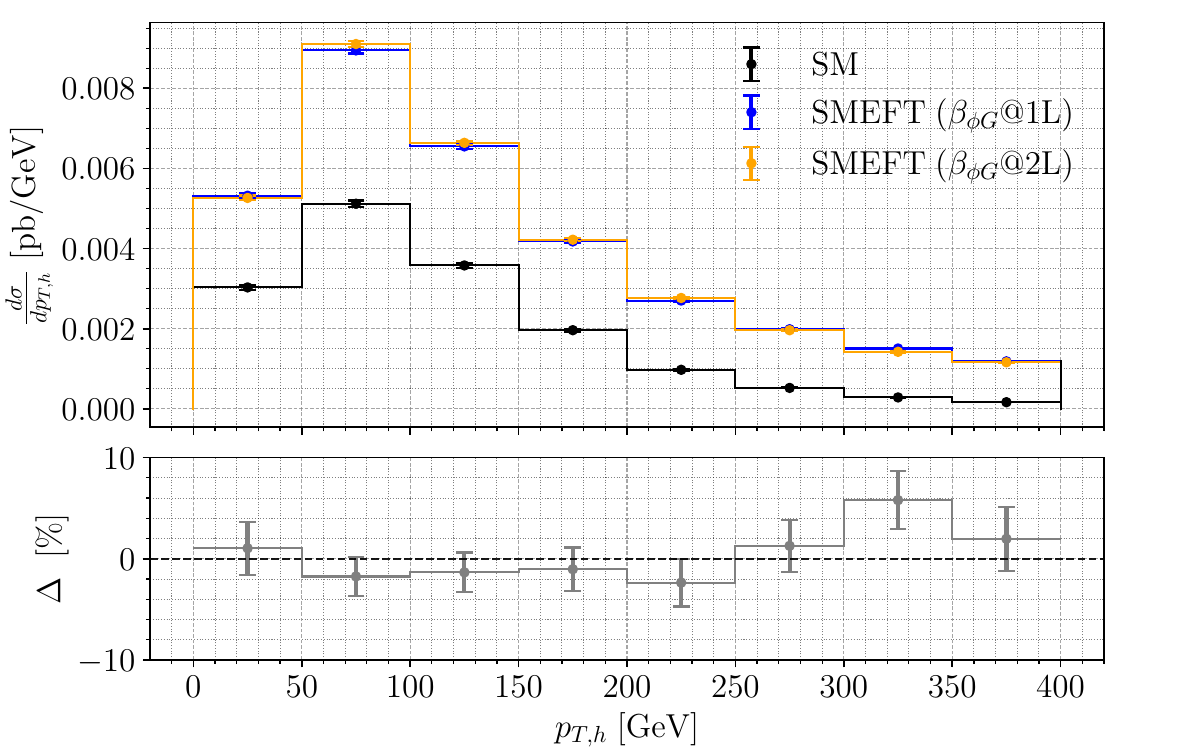}
    \caption{Higgs transverse momentum distribution (upper panel) and percentual difference for each bin between the one- and two-loop running computed as $\Delta \equiv  \left({
\left( \frac{d \sigma }{d p_{T,h}} \right)_{\mathrm{1L}}-\left( \frac{d \sigma }{d p_{T,h}} \right)_{\mathrm{2L}}
} \right)/{\left( \frac{d \sigma }{d p_{T,h}} \right)_{\mathrm{2L}}
}$ (lower panel).}
    \label{fig:dsigma1Lvs2L}
  \end{subfigure}
  \hspace{0.2\linewidth}
  \begin{subfigure}[b]{\linewidth}
    \includegraphics[width=\linewidth]{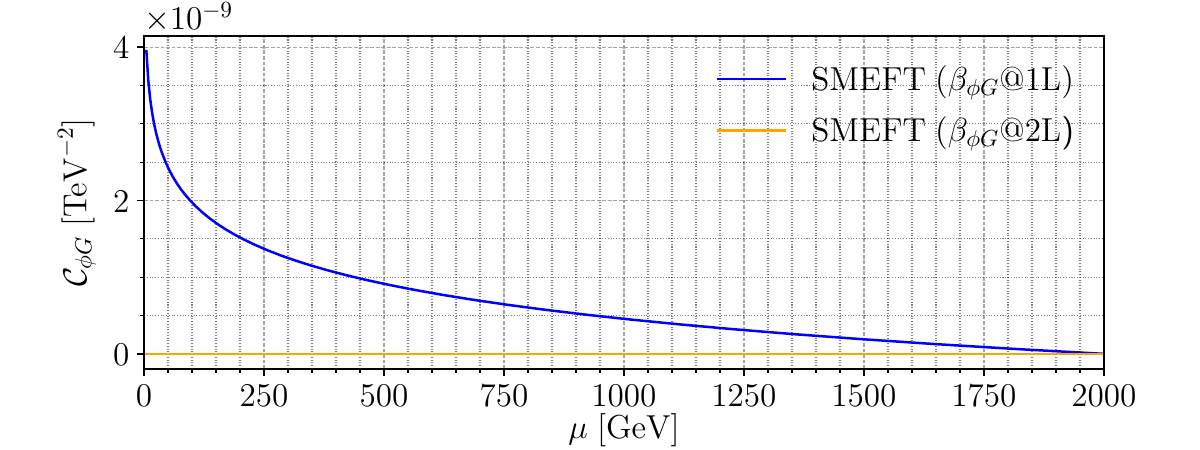}
\caption{Running of $\coeff{\phi G}{}$.}
\label{fig:CHGRunning}
\end{subfigure}
  \caption{Comparison between the one- and two-loop running of $\coeff{\phi G}{}$.}
  \label{fig:1Lvs2L}
\end{figure}

Within this scenario we want to compare the importance of the latter term in comparison to the first one which has been included in previous analyses on running effects, see Refs.~\cite{Grazzini:2018eyk,Deutschmann:2017qum}. We choose $\yuk{\Phi}^2 / M_{\Phi}^2 = 100 / \mathrm{TeV}^2$, which provides SMEFT coefficients which are within the marginalised bounds at $\order{1/\Lambda^2}$ presented in Ref.~\cite{Ethier:2021bye}. We use the first leading-log running (see Eq.~\eqref{eq:approximate}) for all the operators and for the two-loop contributions as the full two-loop RGEs in SMEFT are yet unknown, with some partial results presented in Refs.~\cite{Jenkins:2023bls,Bern:2020ikv}. We report in Fig.~\ref{fig:dsigma1Lvs2L} the results. The two set-ups do not exhibit a large difference (below $10 \%$). The reason is mainly due to the fact that both terms in Eq.~\eqref{eq:RGECphiG} are effectively two-loop order terms (while the SM process arises at tree-level), so their impact is smaller with respect to the one-loop running effects shown in Figs.~\ref{fig:dsigmaQt1} and \ref{fig:dsigmaQt8}. This can also be seen in Fig.~\ref{fig:CHGRunning}.

If the two-loop contribution is not included, $\coeff{\phi G}{}$ increases as the renormalisation scale approaches $0$. When, instead the two-loop contribution of the four-top operators is included, $\coeff{\phi G}{}=0$ at all scales in the leading-logarithmic approximation. We stress that this contribution arises only when NDR is used, being absent in BMHV and that the BMHV and NDR results correspond to each other \textit{only} if the full Eq.~\eqref{eq:RGECphiG} is considered and not just the piece proportional to $\coeff{t G}{}$. 

\paragraph{Scenario 3}: 
In this scenario we want to maximalise the effects of the RG running. Hence
we use the maximal values allowed for the Wilson coefficients by global fits, i.e. adopting the limits from Ref.~\cite{Ethier:2021bye}, at the high energy scale $\Lambda$. We note though that we do not take into account the correlation in the fit, hence our scenario is not compatible with it. However, the scope of this study is not to provide a phenomenologically viable scenario, but it aims to address the importance of running effects.

\begin{figure*}[t]
\centering
\begin{minipage}[b]{0.48\textwidth}
\centering
\includegraphics[width=\linewidth]{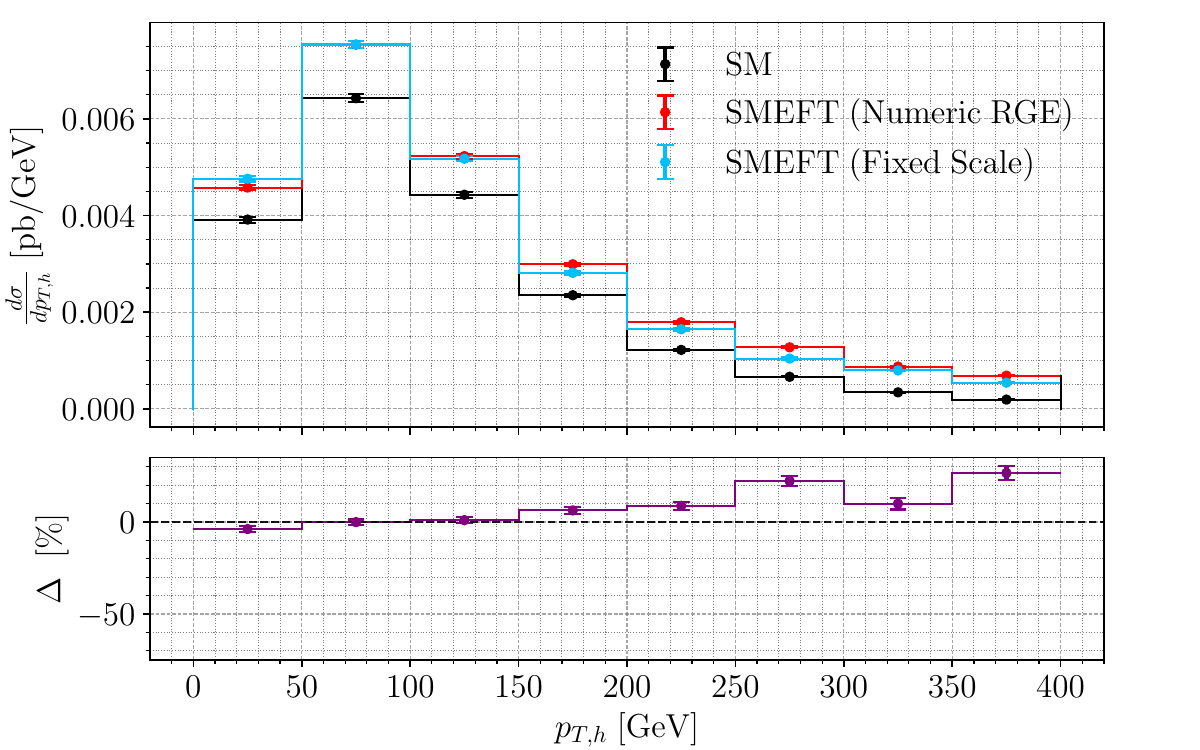}
\caption{Higgs transverse momentum distribution (upper panel)
and percentual difference for each bin between the
dynamical scale and the fixed scale (lower panel). The coefficients are set at $\mu=\Lambda$ following L4, see text for the details.}
    \label{fig:dsigmaMaxL4}
\end{minipage} \hfill
\begin{minipage}[b]{0.48\textwidth}
\centering
\includegraphics[width=\linewidth]{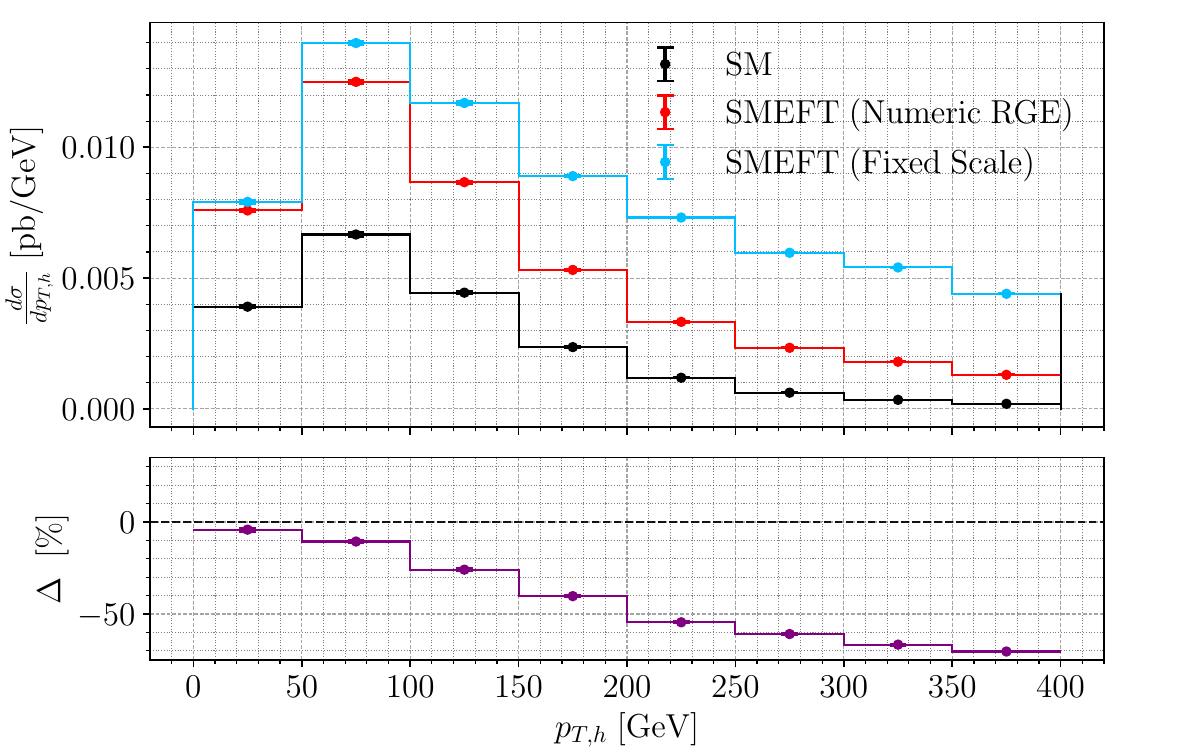}
\caption{Higgs transverse momentum distribution (upper panel)
and percentual difference for each bin between the
dynamical scale and the fixed scale (lower panel). The coefficients are set at $\mu=\Lambda$ following L2, see text for the details.}
\label{fig:dsigmaMaxL2}
\end{minipage}
\end{figure*}

We first rely on the most conservative bounds, namely the marginalised ones obtained at $\order{1/\Lambda^4}$. These bounds provide smaller Wilson coefficients for the operators $\op{Qt}{(1,8)}$ with respect to the previous scenarios. We consider all the four-quark operators (both four-top and two-top-two-light), as well as $\op{\phi \Box }{}$, $\op{\phi D}{}$, $\op{ t \phi}{}$, $\op{tG}{}$, $\op{\phi G}{}$ (translating the basis used in Ref.~\cite{Ethier:2021bye} to ours). This choice of coefficients is referred to as L4.
We set all the operators to one of the extremes of each interval. The result is shown in Fig.~\ref{fig:dsigmaMaxL4}. We observe a difference between the dynamical and the fixed scale of $\sim 25 \%$ in the kinematic tails. 

We now relax the assumptions and consider the $\order{1/\Lambda^2}$ bounds. We set the four-top operators to $100/\Lambda^2$ and the two-top-two-light to $1/\Lambda^2$, with $\Lambda = 2 \, \mathrm{TeV}$.\footnote{To stay within the bounds, we set $\coeff{qq(3)}{3333}(\Lambda) = 80 / \mathrm{TeV}^2$ and $\coeff{uu}{i33i}=0.7/\mathrm{TeV}^2$, being $i=1,2$.} We set the other operators to one of the extremes of the bound interval. This choice of coefficients is referred to as L2. The result is shown in Fig.~\ref{fig:dsigmaMaxL2}. In this case, the effect is larger, up to $70 \%$ in the kinematic tails. 

Finally, we present a comparison between two resummation strategies: numeric solution of the RGEs and first leading-logarithm approximation in Eq.~\eqref{eq:approximate}. The former is more precise, but computationally more expensive (within \texttt{RGESolver}, this method is $\sim 50$ times slower than the approximated solution). 
In order to isolate the contribution from pure SMEFT running, also in this case we use the one-loop SM running for the strong coupling constant $g_s$. The percentual difference in this case is defined as 
\begin{equation}
\Delta \equiv \frac{
\left( \frac{d \sigma }{d p_{T,h}} \right)_{\mathrm{Num}}-\left( \frac{d \sigma }{d p_{T,h}} \right)_{\mathrm{1LL}}
}{\left( \frac{d \sigma }{d p_{T,h}} \right)_{\mathrm{1LL}}
}. \label{eq:percentNumvsLL}
\end{equation}

\begin{figure*}[t]
\centering
\begin{minipage}[b]{0.48\textwidth}
\centering
    \includegraphics[width=\linewidth]{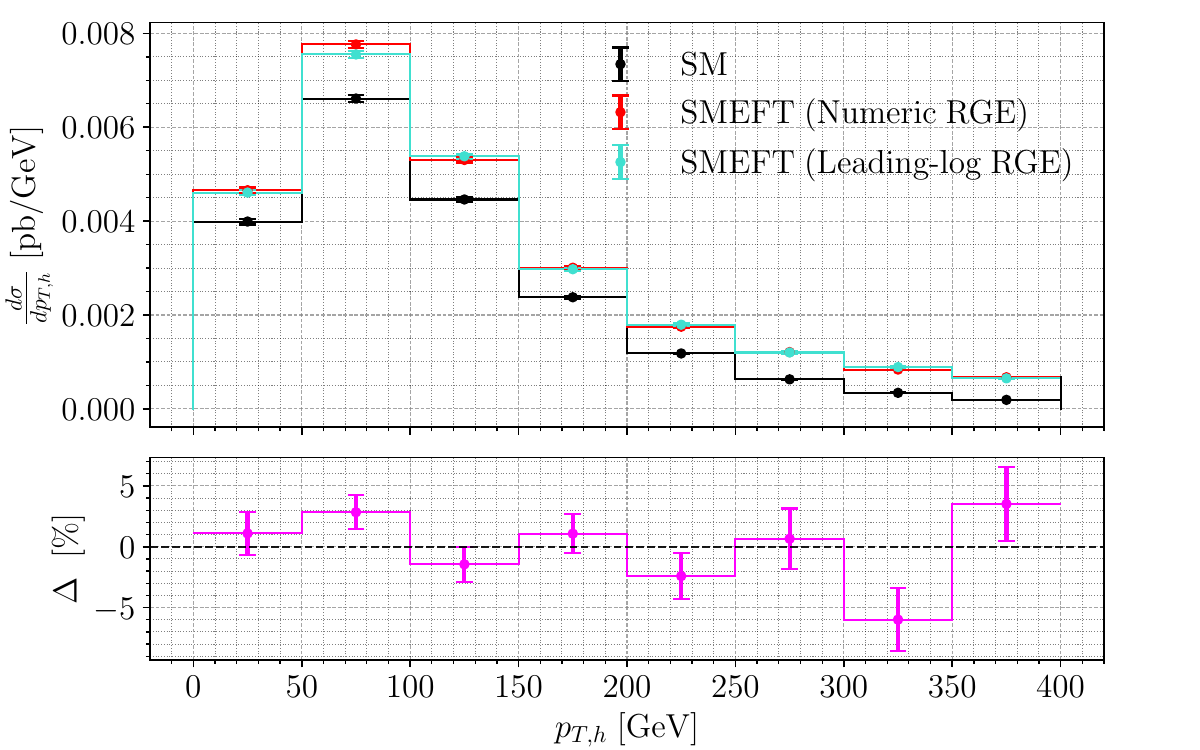}
    \caption{Higgs transverse momentum distribution (upper panel)
and percentual difference for each bin between the
numeric solution of the RGES and the first leading-logarithm approximation as in Eq.~\eqref{eq:percentNumvsLL} (lower panel). The coefficients are set at $\mu=\Lambda$ following L4, see text for the details.}
    \label{fig:dsigmaCompL4}
\end{minipage} \hfill
\begin{minipage}[b]{0.48\textwidth}
 \centering
    \includegraphics[width=\linewidth]{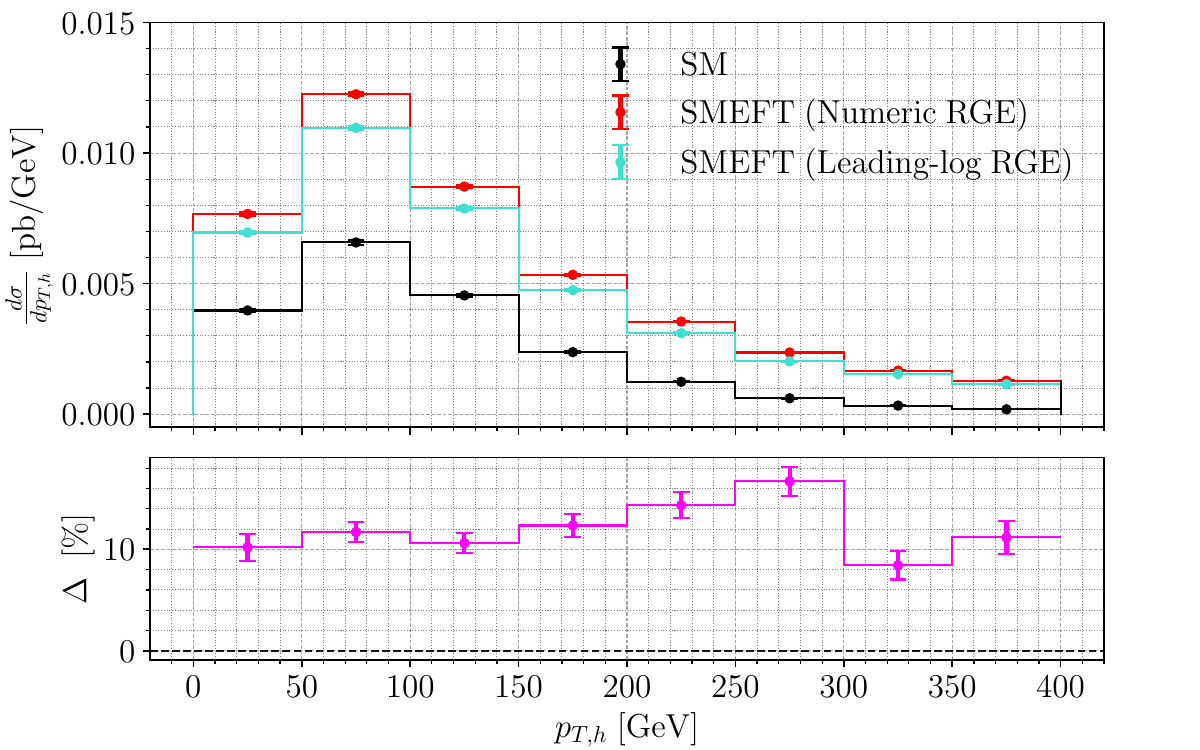}
   \caption{Higgs transverse momentum distribution (upper panel)
and percentual difference for each bin between the
numeric solution of the RGES and the first leading-logarithm approximation as in Eq.~\eqref{eq:percentNumvsLL} (lower panel). The coefficients are set at $\mu=\Lambda$ following L2, see text for the details.}  \label{fig:dsigmaCompL2}
\end{minipage}
\end{figure*}

The results are reported in Fig.~\ref{fig:dsigmaCompL4} (Fig.~\ref{fig:dsigmaCompL2}) for the bounds in scenario L4 (scenario L2). 
In the first case, we do not observe a large deviation between the two methods, oscillating in $[-5 \%,5 \%]$. In the latter case the Wilson coefficients are less constrained and we can see a larger difference, up to $15 \%$.
We thus conclude that, in presence of large Wilson coefficients, the first leading-logarithm solution is not a valid method and to obtain precise results and a numeric solution should be employed. 
\section{Conclusions} \label{sec:conclusions}
We have computed the Higgs transverse momentum distribution for $pp \to t\bar{t}h$ in the SMEFT. We have compared different choices for the renormalisation and factorisation scale: a dynamical scale $\mu_{\mathrm{R}}=\mu_{\mathrm{F}}=H_T/2$ (varying event by event) and a fixed scale $\mu_{\mathrm{R}}=m_t$. Understanding if the latter choice is a valid approximation is crucial, due to the large computational cost of the former. 
We computed the renormalisation group effects at one-loop level numerically using \texttt{RGESolver}. This made it possible to include not only the running of the strong coupling constant as for instance done in Refs.~\cite{Maltoni:2016yxb,Aoude:2022aro,Grazzini:2018eyk, Battaglia:2021nys} but also the contributions proportional to the top quark Yukawa coupling. 

We studied three different scenarios for the Wilson coefficients. In Scenario 1 we analysed individually the effect of the four-top operators $\coeff{Qt}{(1,8)}$. We noted a small difference between the two choices of renormalisation scale, with substantial agreement around $p_{T,h} \sim 100  \,\mathrm{GeV}$. Moreover, we argued that for these operators the running effects proportional to the top Yukawa coupling can be important.

In Scenario 2 we studied the interplay of the same operators with the chromomagnetic operator, in the convenient framework of an UV model involving a heavy scalar $\Phi \sim (8,2)_{1/2}$. In particular, we studied the impact of the two-loop contributions from four-top operators to the Higgs-gluon coupling. Remarkably, these contributions depend on the $\gamma_5$ scheme, as detailed in Ref.~\cite{DiNoi:2023ygk}. In this work, we have shown that their effect is rather small for $pp \to t\bar{t}h$, being a two-loop effect, while the SM matrix element arises at tree-level. However, we stress that this effect can be more relevant for other processes.

Finally, in Scenario 3 we studied the case where the Wilson coefficients assume the maximal values allowed by the bounds presented in Ref.~\cite{Ethier:2021bye}. While this scenario is somewhat not very realistic, at the same time it gives an estimate on how large the running effects can be. We found an effect up to $25 \%$ if we use the L4 scenario where we have set the Wilson coefficient to one of the bounds quoted at $\order{1/\Lambda^4}$ in Ref.~\cite{Ethier:2021bye}. If we use the scenario L2 where we set the four-fermion operator coefficients to $100/\text{TeV}^2$ we observed even bigger effects, up to $70 \, \%$ in the high transverse momentum bins. The importance of renormalisation group effects depends crucially on the size of the Wilson coefficients. In many scenarios where the new physics couples dominantly to top quarks, leading to large coefficients for the four-top operators, such effects should be taken into account. 

Lastly, we compared the numeric solution to the first leading-log approximation. When employing the L4 setting of the Wilson coefficients the difference between the two methods is small, below $5 \%$. When instead the large Wilson coefficients of L2 are used, the two methods show a sizeable deviation from each other, up to $15 \%$. We thus conclude that, when Wilson coefficients are allowed to be large, the approximate solution is not a reliable strategy, calling for a numeric solution of the renormalisation group equations. 

\section*{Acknowledgments}
The Feynman diagrams shown in this work
were drawn with \texttt{TikZ-Feynman}, see Ref.~\cite{Ellis:2016jkw}. SDN would like to thank Gabriele Levati, Laura Reina and Luca Silvestrini for useful discussions.
This project has received funding from the European Union’s Horizon Europe research and innovation programme under the Marie Skłodowska-Curie Staff Exchange  grant agreement No 101086085 – ASYMMETRY. This work is supported in part by the Italian MUR Departments of Excellence grant 2023-2027 "Quantum Frontiers” and by the PNRR CN1-Spoke 2.  SDN also thanks the Lawrence Berkeley
National Laboratory and the Berkeley Center for Theoretical Physics for hospitality.

\bibliography{Biblio1}

\begin{thebibliography}{59}%
\makeatletter
\providecommand \@ifxundefined [1]{%
 \@ifx{#1\undefined}
}%
\providecommand \@ifnum [1]{%
 \ifnum #1\expandafter \@firstoftwo
 \else \expandafter \@secondoftwo
 \fi
}%
\providecommand \@ifx [1]{%
 \ifx #1\expandafter \@firstoftwo
 \else \expandafter \@secondoftwo
 \fi
}%
\providecommand \natexlab [1]{#1}%
\providecommand \enquote  [1]{``#1''}%
\providecommand \bibnamefont  [1]{#1}%
\providecommand \bibfnamefont [1]{#1}%
\providecommand \citenamefont [1]{#1}%
\providecommand \href@noop [0]{\@secondoftwo}%
\providecommand \href [0]{\begingroup \@sanitize@url \@href}%
\providecommand \@href[1]{\@@startlink{#1}\@@href}%
\providecommand \@@href[1]{\endgroup#1\@@endlink}%
\providecommand \@sanitize@url [0]{\catcode `\\12\catcode `\$12\catcode
  `\&12\catcode `\#12\catcode `\^12\catcode `\_12\catcode `\%12\relax}%
\providecommand \@@startlink[1]{}%
\providecommand \@@endlink[0]{}%
\providecommand \url  [0]{\begingroup\@sanitize@url \@url }%
\providecommand \@url [1]{\endgroup\@href {#1}{\urlprefix }}%
\providecommand \urlprefix  [0]{URL }%
\providecommand \Eprint [0]{\href }%
\providecommand \doibase [0]{http://dx.doi.org/}%
\providecommand \selectlanguage [0]{\@gobble}%
\providecommand \bibinfo  [0]{\@secondoftwo}%
\providecommand \bibfield  [0]{\@secondoftwo}%
\providecommand \translation [1]{[#1]}%
\providecommand \BibitemOpen [0]{}%
\providecommand \bibitemStop [0]{}%
\providecommand \bibitemNoStop [0]{.\EOS\space}%
\providecommand \EOS [0]{\spacefactor3000\relax}%
\providecommand \BibitemShut  [1]{\csname bibitem#1\endcsname}%
\let\auto@bib@innerbib\@empty
\bibitem [{\citenamefont {Aad}\ \emph {et~al.}(2012)\citenamefont {Aad} \emph
  {et~al.}}]{HiggsATLAS}%
  \BibitemOpen
  \bibfield  {author} {\bibinfo {author} {\bibfnamefont {G.}~\bibnamefont
  {Aad}} \emph {et~al.} (\bibinfo {collaboration} {ATLAS}),\ }\href {\doibase
  10.1016/j.physletb.2012.08.020} {\bibfield  {journal} {\bibinfo  {journal}
  {Phys. Lett. B}\ }\textbf {\bibinfo {volume} {716}},\ \bibinfo {pages} {1}
  (\bibinfo {year} {2012})},\ \Eprint {http://arxiv.org/abs/1207.7214}
  {arXiv:1207.7214 [hep-ex]} \BibitemShut {NoStop}%
\bibitem [{\citenamefont {Chatrchyan}\ \emph {et~al.}(2012)\citenamefont
  {Chatrchyan} \emph {et~al.}}]{HiggsCMS}%
  \BibitemOpen
  \bibfield  {author} {\bibinfo {author} {\bibfnamefont {S.}~\bibnamefont
  {Chatrchyan}} \emph {et~al.} (\bibinfo {collaboration} {CMS}),\ }\href
  {\doibase 10.1016/j.physletb.2012.08.021} {\bibfield  {journal} {\bibinfo
  {journal} {Phys. Lett. B}\ }\textbf {\bibinfo {volume} {716}},\ \bibinfo
  {pages} {30} (\bibinfo {year} {2012})},\ \Eprint
  {http://arxiv.org/abs/1207.7235} {arXiv:1207.7235 [hep-ex]} \BibitemShut
  {NoStop}%
\bibitem [{\citenamefont {Buchmuller}\ and\ \citenamefont
  {Wyler}(1986)}]{Buchmuller:1985jz}%
  \BibitemOpen
  \bibfield  {author} {\bibinfo {author} {\bibfnamefont {W.}~\bibnamefont
  {Buchmuller}}\ and\ \bibinfo {author} {\bibfnamefont {D.}~\bibnamefont
  {Wyler}},\ }\href {\doibase 10.1016/0550-3213(86)90262-2} {\bibfield
  {journal} {\bibinfo  {journal} {Nucl. Phys. B}\ }\textbf {\bibinfo {volume}
  {268}},\ \bibinfo {pages} {621} (\bibinfo {year} {1986})}\BibitemShut
  {NoStop}%
\bibitem [{\citenamefont {Grzadkowski}\ \emph {et~al.}(2010)\citenamefont
  {Grzadkowski}, \citenamefont {Iskrzynski}, \citenamefont {Misiak},\ and\
  \citenamefont {Rosiek}}]{dim6smeft}%
  \BibitemOpen
  \bibfield  {author} {\bibinfo {author} {\bibfnamefont {B.}~\bibnamefont
  {Grzadkowski}}, \bibinfo {author} {\bibfnamefont {M.}~\bibnamefont
  {Iskrzynski}}, \bibinfo {author} {\bibfnamefont {M.}~\bibnamefont {Misiak}},
  \ and\ \bibinfo {author} {\bibfnamefont {J.}~\bibnamefont {Rosiek}},\ }\href
  {\doibase 10.1007/JHEP10(2010)085} {\bibfield  {journal} {\bibinfo  {journal}
  {JHEP}\ }\textbf {\bibinfo {volume} {10}},\ \bibinfo {pages} {085} (\bibinfo
  {year} {2010})},\ \Eprint {http://arxiv.org/abs/1008.4884} {arXiv:1008.4884
  [hep-ph]} \BibitemShut {NoStop}%
\bibitem [{\citenamefont {Giudice}\ \emph {et~al.}(2007)\citenamefont
  {Giudice}, \citenamefont {Grojean}, \citenamefont {Pomarol},\ and\
  \citenamefont {Rattazzi}}]{Giudice:2007fh}%
  \BibitemOpen
  \bibfield  {author} {\bibinfo {author} {\bibfnamefont {G.~F.}\ \bibnamefont
  {Giudice}}, \bibinfo {author} {\bibfnamefont {C.}~\bibnamefont {Grojean}},
  \bibinfo {author} {\bibfnamefont {A.}~\bibnamefont {Pomarol}}, \ and\
  \bibinfo {author} {\bibfnamefont {R.}~\bibnamefont {Rattazzi}},\ }\href
  {\doibase 10.1088/1126-6708/2007/06/045} {\bibfield  {journal} {\bibinfo
  {journal} {JHEP}\ }\textbf {\bibinfo {volume} {06}},\ \bibinfo {pages} {045}
  (\bibinfo {year} {2007})},\ \Eprint {http://arxiv.org/abs/hep-ph/0703164}
  {arXiv:hep-ph/0703164} \BibitemShut {NoStop}%
\bibitem [{\citenamefont {Contino}\ \emph {et~al.}(2013)\citenamefont
  {Contino}, \citenamefont {Ghezzi}, \citenamefont {Grojean}, \citenamefont
  {Muhlleitner},\ and\ \citenamefont {Spira}}]{Contino:2013kra}%
  \BibitemOpen
  \bibfield  {author} {\bibinfo {author} {\bibfnamefont {R.}~\bibnamefont
  {Contino}}, \bibinfo {author} {\bibfnamefont {M.}~\bibnamefont {Ghezzi}},
  \bibinfo {author} {\bibfnamefont {C.}~\bibnamefont {Grojean}}, \bibinfo
  {author} {\bibfnamefont {M.}~\bibnamefont {Muhlleitner}}, \ and\ \bibinfo
  {author} {\bibfnamefont {M.}~\bibnamefont {Spira}},\ }\href {\doibase
  10.1007/JHEP07(2013)035} {\bibfield  {journal} {\bibinfo  {journal} {JHEP}\
  }\textbf {\bibinfo {volume} {07}},\ \bibinfo {pages} {035} (\bibinfo {year}
  {2013})},\ \Eprint {http://arxiv.org/abs/1303.3876} {arXiv:1303.3876
  [hep-ph]} \BibitemShut {NoStop}%
\bibitem [{\citenamefont {Elias-Mir\'o}\ \emph {et~al.}(2014)\citenamefont
  {Elias-Mir\'o}, \citenamefont {Grojean}, \citenamefont {Gupta},\ and\
  \citenamefont {Marzocca}}]{Elias-Miro:2013eta}%
  \BibitemOpen
  \bibfield  {author} {\bibinfo {author} {\bibfnamefont {J.}~\bibnamefont
  {Elias-Mir\'o}}, \bibinfo {author} {\bibfnamefont {C.}~\bibnamefont
  {Grojean}}, \bibinfo {author} {\bibfnamefont {R.~S.}\ \bibnamefont {Gupta}},
  \ and\ \bibinfo {author} {\bibfnamefont {D.}~\bibnamefont {Marzocca}},\
  }\href {\doibase 10.1007/JHEP05(2014)019} {\bibfield  {journal} {\bibinfo
  {journal} {JHEP}\ }\textbf {\bibinfo {volume} {05}},\ \bibinfo {pages} {019}
  (\bibinfo {year} {2014})},\ \Eprint {http://arxiv.org/abs/1312.2928}
  {arXiv:1312.2928 [hep-ph]} \BibitemShut {NoStop}%
\bibitem [{\citenamefont {Manohar}(2020)}]{man}%
  \BibitemOpen
  \bibfield  {author} {\bibinfo {author} {\bibfnamefont {A.~V.}\ \bibnamefont
  {Manohar}},\ }\href {\doibase 10.1093/oso/9780198855743.003.0002} {\bibfield
  {journal} {\bibinfo  {journal} {Les Houches Lect. Notes}\ }\textbf {\bibinfo
  {volume} {108}} (\bibinfo {year} {2020}),\
  10.1093/oso/9780198855743.003.0002},\ \Eprint
  {http://arxiv.org/abs/1804.05863} {arXiv:1804.05863 [hep-ph]} \BibitemShut
  {NoStop}%
\bibitem [{\citenamefont {Jenkins}\ \emph {et~al.}(2013)\citenamefont
  {Jenkins}, \citenamefont {Manohar},\ and\ \citenamefont {Trott}}]{rge1}%
  \BibitemOpen
  \bibfield  {author} {\bibinfo {author} {\bibfnamefont {E.~E.}\ \bibnamefont
  {Jenkins}}, \bibinfo {author} {\bibfnamefont {A.~V.}\ \bibnamefont
  {Manohar}}, \ and\ \bibinfo {author} {\bibfnamefont {M.}~\bibnamefont
  {Trott}},\ }\href {\doibase 10.1007/JHEP10(2013)087} {\bibfield  {journal}
  {\bibinfo  {journal} {JHEP}\ }\textbf {\bibinfo {volume} {10}},\ \bibinfo
  {pages} {087} (\bibinfo {year} {2013})},\ \Eprint
  {http://arxiv.org/abs/1308.2627v4} {arXiv:1308.2627v4 [hep-ph]} \BibitemShut
  {NoStop}%
\bibitem [{\citenamefont {Jenkins}\ \emph {et~al.}(2014)\citenamefont
  {Jenkins}, \citenamefont {Manohar},\ and\ \citenamefont {Trott}}]{rge2}%
  \BibitemOpen
  \bibfield  {author} {\bibinfo {author} {\bibfnamefont {E.~E.}\ \bibnamefont
  {Jenkins}}, \bibinfo {author} {\bibfnamefont {A.~V.}\ \bibnamefont
  {Manohar}}, \ and\ \bibinfo {author} {\bibfnamefont {M.}~\bibnamefont
  {Trott}},\ }\href {\doibase 10.1007/JHEP01(2014)035} {\bibfield  {journal}
  {\bibinfo  {journal} {JHEP}\ }\textbf {\bibinfo {volume} {01}},\ \bibinfo
  {pages} {035} (\bibinfo {year} {2014})},\ \Eprint
  {http://arxiv.org/abs/1310.4838v3} {arXiv:1310.4838v3 [hep-ph]} \BibitemShut
  {NoStop}%
\bibitem [{\citenamefont {Alonso}\ \emph {et~al.}(2014)\citenamefont {Alonso},
  \citenamefont {Jenkins}, \citenamefont {Manohar},\ and\ \citenamefont
  {Trott}}]{rge3}%
  \BibitemOpen
  \bibfield  {author} {\bibinfo {author} {\bibfnamefont {R.}~\bibnamefont
  {Alonso}}, \bibinfo {author} {\bibfnamefont {E.~E.}\ \bibnamefont {Jenkins}},
  \bibinfo {author} {\bibfnamefont {A.~V.}\ \bibnamefont {Manohar}}, \ and\
  \bibinfo {author} {\bibfnamefont {M.}~\bibnamefont {Trott}},\ }\href
  {\doibase 10.1007/JHEP04(2014)159} {\bibfield  {journal} {\bibinfo  {journal}
  {JHEP}\ }\textbf {\bibinfo {volume} {04}},\ \bibinfo {pages} {159} (\bibinfo
  {year} {2014})},\ \Eprint {http://arxiv.org/abs/1312.2014v4}
  {arXiv:1312.2014v4 [hep-ph]} \BibitemShut {NoStop}%
\bibitem [{\citenamefont {Criado}(2018)}]{Criado:2017khh}%
  \BibitemOpen
  \bibfield  {author} {\bibinfo {author} {\bibfnamefont {J.~C.}\ \bibnamefont
  {Criado}},\ }\href {\doibase 10.1016/j.cpc.2018.02.016} {\bibfield  {journal}
  {\bibinfo  {journal} {Comput. Phys. Commun.}\ }\textbf {\bibinfo {volume}
  {227}},\ \bibinfo {pages} {42} (\bibinfo {year} {2018})},\ \Eprint
  {http://arxiv.org/abs/1710.06445} {arXiv:1710.06445 [hep-ph]} \BibitemShut
  {NoStop}%
\bibitem [{\citenamefont {Das~Bakshi}\ \emph {et~al.}(2019)\citenamefont
  {Das~Bakshi}, \citenamefont {Chakrabortty},\ and\ \citenamefont
  {Patra}}]{DasBakshi:2018vni}%
  \BibitemOpen
  \bibfield  {author} {\bibinfo {author} {\bibfnamefont {S.}~\bibnamefont
  {Das~Bakshi}}, \bibinfo {author} {\bibfnamefont {J.}~\bibnamefont
  {Chakrabortty}}, \ and\ \bibinfo {author} {\bibfnamefont {S.~K.}\
  \bibnamefont {Patra}},\ }\href {\doibase 10.1140/epjc/s10052-018-6444-2}
  {\bibfield  {journal} {\bibinfo  {journal} {Eur. Phys. J. C}\ }\textbf
  {\bibinfo {volume} {79}},\ \bibinfo {pages} {21} (\bibinfo {year} {2019})},\
  \Eprint {http://arxiv.org/abs/1808.04403} {arXiv:1808.04403 [hep-ph]}
  \BibitemShut {NoStop}%
\bibitem [{\citenamefont {Cohen}\ \emph {et~al.}(2021)\citenamefont {Cohen},
  \citenamefont {Lu},\ and\ \citenamefont {Zhang}}]{Cohen:2020qvb}%
  \BibitemOpen
  \bibfield  {author} {\bibinfo {author} {\bibfnamefont {T.}~\bibnamefont
  {Cohen}}, \bibinfo {author} {\bibfnamefont {X.}~\bibnamefont {Lu}}, \ and\
  \bibinfo {author} {\bibfnamefont {Z.}~\bibnamefont {Zhang}},\ }\href
  {\doibase 10.21468/SciPostPhys.10.5.098} {\bibfield  {journal} {\bibinfo
  {journal} {SciPost Phys.}\ }\textbf {\bibinfo {volume} {10}},\ \bibinfo
  {pages} {098} (\bibinfo {year} {2021})},\ \Eprint
  {http://arxiv.org/abs/2012.07851} {arXiv:2012.07851 [hep-ph]} \BibitemShut
  {NoStop}%
\bibitem [{\citenamefont {Fuentes-Mart\'\i{}n}\ \emph
  {et~al.}(2022)\citenamefont {Fuentes-Mart\'\i{}n}, \citenamefont {K\"onig},
  \citenamefont {Pag\`es}, \citenamefont {Thomsen},\ and\ \citenamefont
  {Wilsch}}]{Fuentes-Martin:2022jrf}%
  \BibitemOpen
  \bibfield  {author} {\bibinfo {author} {\bibfnamefont {J.}~\bibnamefont
  {Fuentes-Mart\'\i{}n}}, \bibinfo {author} {\bibfnamefont {M.}~\bibnamefont
  {K\"onig}}, \bibinfo {author} {\bibfnamefont {J.}~\bibnamefont {Pag\`es}},
  \bibinfo {author} {\bibfnamefont {A.~E.}\ \bibnamefont {Thomsen}}, \ and\
  \bibinfo {author} {\bibfnamefont {F.}~\bibnamefont {Wilsch}},\ }\href@noop {}
  {\  (\bibinfo {year} {2022})},\ \Eprint {http://arxiv.org/abs/2212.04510}
  {arXiv:2212.04510 [hep-ph]} \BibitemShut {NoStop}%
\bibitem [{\citenamefont {Guedes}\ \emph {et~al.}(2023)\citenamefont {Guedes},
  \citenamefont {Olgoso},\ and\ \citenamefont {Santiago}}]{Guedes:2023azv}%
  \BibitemOpen
  \bibfield  {author} {\bibinfo {author} {\bibfnamefont {G.}~\bibnamefont
  {Guedes}}, \bibinfo {author} {\bibfnamefont {P.}~\bibnamefont {Olgoso}}, \
  and\ \bibinfo {author} {\bibfnamefont {J.}~\bibnamefont {Santiago}},\ }\href
  {\doibase 10.21468/SciPostPhys.15.4.143} {\bibfield  {journal} {\bibinfo
  {journal} {SciPost Phys.}\ }\textbf {\bibinfo {volume} {15}},\ \bibinfo
  {pages} {143} (\bibinfo {year} {2023})},\ \Eprint
  {http://arxiv.org/abs/2303.16965} {arXiv:2303.16965 [hep-ph]} \BibitemShut
  {NoStop}%
\bibitem [{\citenamefont {Maltoni}\ \emph {et~al.}(2016)\citenamefont
  {Maltoni}, \citenamefont {Vryonidou},\ and\ \citenamefont
  {Zhang}}]{Maltoni:2016yxb}%
  \BibitemOpen
  \bibfield  {author} {\bibinfo {author} {\bibfnamefont {F.}~\bibnamefont
  {Maltoni}}, \bibinfo {author} {\bibfnamefont {E.}~\bibnamefont {Vryonidou}},
  \ and\ \bibinfo {author} {\bibfnamefont {C.}~\bibnamefont {Zhang}},\ }\href
  {\doibase 10.1007/JHEP10(2016)123} {\bibfield  {journal} {\bibinfo  {journal}
  {JHEP}\ }\textbf {\bibinfo {volume} {10}},\ \bibinfo {pages} {123} (\bibinfo
  {year} {2016})},\ \Eprint {http://arxiv.org/abs/1607.05330} {arXiv:1607.05330
  [hep-ph]} \BibitemShut {NoStop}%
\bibitem [{\citenamefont {Aoude}\ \emph {et~al.}(2023)\citenamefont {Aoude},
  \citenamefont {Maltoni}, \citenamefont {Mattelaer}, \citenamefont {Severi},\
  and\ \citenamefont {Vryonidou}}]{Aoude:2022aro}%
  \BibitemOpen
  \bibfield  {author} {\bibinfo {author} {\bibfnamefont {R.}~\bibnamefont
  {Aoude}}, \bibinfo {author} {\bibfnamefont {F.}~\bibnamefont {Maltoni}},
  \bibinfo {author} {\bibfnamefont {O.}~\bibnamefont {Mattelaer}}, \bibinfo
  {author} {\bibfnamefont {C.}~\bibnamefont {Severi}}, \ and\ \bibinfo {author}
  {\bibfnamefont {E.}~\bibnamefont {Vryonidou}},\ }\href {\doibase
  10.1007/JHEP09(2023)191} {\bibfield  {journal} {\bibinfo  {journal} {JHEP}\
  }\textbf {\bibinfo {volume} {09}},\ \bibinfo {pages} {191} (\bibinfo {year}
  {2023})},\ \Eprint {http://arxiv.org/abs/2212.05067} {arXiv:2212.05067
  [hep-ph]} \BibitemShut {NoStop}%
\bibitem [{\citenamefont {Grazzini}\ \emph {et~al.}(2018)\citenamefont
  {Grazzini}, \citenamefont {Ilnicka},\ and\ \citenamefont
  {Spira}}]{Grazzini:2018eyk}%
  \BibitemOpen
  \bibfield  {author} {\bibinfo {author} {\bibfnamefont {M.}~\bibnamefont
  {Grazzini}}, \bibinfo {author} {\bibfnamefont {A.}~\bibnamefont {Ilnicka}}, \
  and\ \bibinfo {author} {\bibfnamefont {M.}~\bibnamefont {Spira}},\ }\href
  {\doibase 10.1140/epjc/s10052-018-6261-7} {\bibfield  {journal} {\bibinfo
  {journal} {Eur. Phys. J. C}\ }\textbf {\bibinfo {volume} {78}},\ \bibinfo
  {pages} {808} (\bibinfo {year} {2018})},\ \Eprint
  {http://arxiv.org/abs/1806.08832} {arXiv:1806.08832 [hep-ph]} \BibitemShut
  {NoStop}%
\bibitem [{\citenamefont {Battaglia}\ \emph {et~al.}(2021)\citenamefont
  {Battaglia}, \citenamefont {Grazzini}, \citenamefont {Spira},\ and\
  \citenamefont {Wiesemann}}]{Battaglia:2021nys}%
  \BibitemOpen
  \bibfield  {author} {\bibinfo {author} {\bibfnamefont {M.}~\bibnamefont
  {Battaglia}}, \bibinfo {author} {\bibfnamefont {M.}~\bibnamefont {Grazzini}},
  \bibinfo {author} {\bibfnamefont {M.}~\bibnamefont {Spira}}, \ and\ \bibinfo
  {author} {\bibfnamefont {M.}~\bibnamefont {Wiesemann}},\ }\href {\doibase
  10.1007/JHEP11(2021)173} {\bibfield  {journal} {\bibinfo  {journal} {JHEP}\
  }\textbf {\bibinfo {volume} {11}},\ \bibinfo {pages} {173} (\bibinfo {year}
  {2021})},\ \Eprint {http://arxiv.org/abs/2109.02987} {arXiv:2109.02987
  [hep-ph]} \BibitemShut {NoStop}%
\bibitem [{\citenamefont {Ethier}\ \emph {et~al.}(2021)\citenamefont {Ethier},
  \citenamefont {Magni}, \citenamefont {Maltoni}, \citenamefont {Mantani},
  \citenamefont {Nocera}, \citenamefont {Rojo}, \citenamefont {Slade},
  \citenamefont {Vryonidou},\ and\ \citenamefont {Zhang}}]{Ethier:2021bye}%
  \BibitemOpen
  \bibfield  {author} {\bibinfo {author} {\bibfnamefont {J.~J.}\ \bibnamefont
  {Ethier}}, \bibinfo {author} {\bibfnamefont {G.}~\bibnamefont {Magni}},
  \bibinfo {author} {\bibfnamefont {F.}~\bibnamefont {Maltoni}}, \bibinfo
  {author} {\bibfnamefont {L.}~\bibnamefont {Mantani}}, \bibinfo {author}
  {\bibfnamefont {E.~R.}\ \bibnamefont {Nocera}}, \bibinfo {author}
  {\bibfnamefont {J.}~\bibnamefont {Rojo}}, \bibinfo {author} {\bibfnamefont
  {E.}~\bibnamefont {Slade}}, \bibinfo {author} {\bibfnamefont
  {E.}~\bibnamefont {Vryonidou}}, \ and\ \bibinfo {author} {\bibfnamefont
  {C.}~\bibnamefont {Zhang}} (\bibinfo {collaboration} {SMEFiT}),\ }\href
  {\doibase 10.1007/JHEP11(2021)089} {\bibfield  {journal} {\bibinfo  {journal}
  {JHEP}\ }\textbf {\bibinfo {volume} {11}},\ \bibinfo {pages} {089} (\bibinfo
  {year} {2021})},\ \Eprint {http://arxiv.org/abs/2105.00006} {arXiv:2105.00006
  [hep-ph]} \BibitemShut {NoStop}%
\bibitem [{\citenamefont {Dedes}\ \emph {et~al.}(2017)\citenamefont {Dedes},
  \citenamefont {Materkowska}, \citenamefont {Paraskevas}, \citenamefont
  {Rosiek},\ and\ \citenamefont {Suxho}}]{FeynRules}%
  \BibitemOpen
  \bibfield  {author} {\bibinfo {author} {\bibfnamefont {A.}~\bibnamefont
  {Dedes}}, \bibinfo {author} {\bibfnamefont {W.}~\bibnamefont {Materkowska}},
  \bibinfo {author} {\bibfnamefont {M.}~\bibnamefont {Paraskevas}}, \bibinfo
  {author} {\bibfnamefont {J.}~\bibnamefont {Rosiek}}, \ and\ \bibinfo {author}
  {\bibfnamefont {K.}~\bibnamefont {Suxho}},\ }\href {\doibase
  10.1007/JHEP06(2017)143} {\bibfield  {journal} {\bibinfo  {journal} {JHEP}\
  }\textbf {\bibinfo {volume} {06}},\ \bibinfo {pages} {143} (\bibinfo {year}
  {2017})},\ \Eprint {http://arxiv.org/abs/1704.03888} {arXiv:1704.03888
  [hep-ph]} \BibitemShut {NoStop}%
\bibitem [{\citenamefont {Jenkins}\ \emph {et~al.}(2018)\citenamefont
  {Jenkins}, \citenamefont {Manohar},\ and\ \citenamefont
  {Stoffer}}]{Jenkins:2017jig}%
  \BibitemOpen
  \bibfield  {author} {\bibinfo {author} {\bibfnamefont {E.~E.}\ \bibnamefont
  {Jenkins}}, \bibinfo {author} {\bibfnamefont {A.~V.}\ \bibnamefont
  {Manohar}}, \ and\ \bibinfo {author} {\bibfnamefont {P.}~\bibnamefont
  {Stoffer}},\ }\href {\doibase 10.1007/JHEP03(2018)016} {\bibfield  {journal}
  {\bibinfo  {journal} {JHEP}\ }\textbf {\bibinfo {volume} {03}},\ \bibinfo
  {pages} {016} (\bibinfo {year} {2018})},\ \Eprint
  {http://arxiv.org/abs/1709.04486} {arXiv:1709.04486 [hep-ph]} \BibitemShut
  {NoStop}%
\bibitem [{\citenamefont {Chien}\ \emph {et~al.}(2016)\citenamefont {Chien},
  \citenamefont {Cirigliano}, \citenamefont {Dekens}, \citenamefont
  {de~Vries},\ and\ \citenamefont {Mereghetti}}]{Chien:2015xha}%
  \BibitemOpen
  \bibfield  {author} {\bibinfo {author} {\bibfnamefont {Y.~T.}\ \bibnamefont
  {Chien}}, \bibinfo {author} {\bibfnamefont {V.}~\bibnamefont {Cirigliano}},
  \bibinfo {author} {\bibfnamefont {W.}~\bibnamefont {Dekens}}, \bibinfo
  {author} {\bibfnamefont {J.}~\bibnamefont {de~Vries}}, \ and\ \bibinfo
  {author} {\bibfnamefont {E.}~\bibnamefont {Mereghetti}},\ }\href {\doibase
  10.1007/JHEP02(2016)011} {\bibfield  {journal} {\bibinfo  {journal} {JHEP}\
  }\textbf {\bibinfo {volume} {02}},\ \bibinfo {pages} {011} (\bibinfo {year}
  {2016})},\ \Eprint {http://arxiv.org/abs/1510.00725} {arXiv:1510.00725
  [hep-ph]} \BibitemShut {NoStop}%
\bibitem [{\citenamefont {Cirigliano}\ \emph {et~al.}(2016)\citenamefont
  {Cirigliano}, \citenamefont {Dekens}, \citenamefont {de~Vries},\ and\
  \citenamefont {Mereghetti}}]{Cirigliano:2016njn}%
  \BibitemOpen
  \bibfield  {author} {\bibinfo {author} {\bibfnamefont {V.}~\bibnamefont
  {Cirigliano}}, \bibinfo {author} {\bibfnamefont {W.}~\bibnamefont {Dekens}},
  \bibinfo {author} {\bibfnamefont {J.}~\bibnamefont {de~Vries}}, \ and\
  \bibinfo {author} {\bibfnamefont {E.}~\bibnamefont {Mereghetti}},\ }\href
  {\doibase 10.1103/PhysRevD.94.016002} {\bibfield  {journal} {\bibinfo
  {journal} {Phys. Rev. D}\ }\textbf {\bibinfo {volume} {94}},\ \bibinfo
  {pages} {016002} (\bibinfo {year} {2016})},\ \Eprint
  {http://arxiv.org/abs/1603.03049} {arXiv:1603.03049 [hep-ph]} \BibitemShut
  {NoStop}%
\bibitem [{\citenamefont {Bahl}\ \emph {et~al.}(2020)\citenamefont {Bahl},
  \citenamefont {Bechtle}, \citenamefont {Heinemeyer}, \citenamefont {Katzy},
  \citenamefont {Klingl}, \citenamefont {Peters}, \citenamefont {Saimpert},
  \citenamefont {Stefaniak},\ and\ \citenamefont {Weiglein}}]{Bahl:2020wee}%
  \BibitemOpen
  \bibfield  {author} {\bibinfo {author} {\bibfnamefont {H.}~\bibnamefont
  {Bahl}}, \bibinfo {author} {\bibfnamefont {P.}~\bibnamefont {Bechtle}},
  \bibinfo {author} {\bibfnamefont {S.}~\bibnamefont {Heinemeyer}}, \bibinfo
  {author} {\bibfnamefont {J.}~\bibnamefont {Katzy}}, \bibinfo {author}
  {\bibfnamefont {T.}~\bibnamefont {Klingl}}, \bibinfo {author} {\bibfnamefont
  {K.}~\bibnamefont {Peters}}, \bibinfo {author} {\bibfnamefont
  {M.}~\bibnamefont {Saimpert}}, \bibinfo {author} {\bibfnamefont
  {T.}~\bibnamefont {Stefaniak}}, \ and\ \bibinfo {author} {\bibfnamefont
  {G.}~\bibnamefont {Weiglein}},\ }\href {\doibase 10.1007/JHEP11(2020)127}
  {\bibfield  {journal} {\bibinfo  {journal} {JHEP}\ }\textbf {\bibinfo
  {volume} {11}},\ \bibinfo {pages} {127} (\bibinfo {year} {2020})},\ \Eprint
  {http://arxiv.org/abs/2007.08542} {arXiv:2007.08542 [hep-ph]} \BibitemShut
  {NoStop}%
\bibitem [{\citenamefont {Martini}\ \emph {et~al.}(2021)\citenamefont
  {Martini}, \citenamefont {Pan}, \citenamefont {Schulze},\ and\ \citenamefont
  {Xiao}}]{Martini:2021uey}%
  \BibitemOpen
  \bibfield  {author} {\bibinfo {author} {\bibfnamefont {T.}~\bibnamefont
  {Martini}}, \bibinfo {author} {\bibfnamefont {R.-Q.}\ \bibnamefont {Pan}},
  \bibinfo {author} {\bibfnamefont {M.}~\bibnamefont {Schulze}}, \ and\
  \bibinfo {author} {\bibfnamefont {M.}~\bibnamefont {Xiao}},\ }\href {\doibase
  10.1103/PhysRevD.104.055045} {\bibfield  {journal} {\bibinfo  {journal}
  {Phys. Rev. D}\ }\textbf {\bibinfo {volume} {104}},\ \bibinfo {pages}
  {055045} (\bibinfo {year} {2021})},\ \Eprint
  {http://arxiv.org/abs/2104.04277} {arXiv:2104.04277 [hep-ph]} \BibitemShut
  {NoStop}%
\bibitem [{\citenamefont {Bahl}\ and\ \citenamefont
  {Brass}(2022)}]{Bahl:2021dnc}%
  \BibitemOpen
  \bibfield  {author} {\bibinfo {author} {\bibfnamefont {H.}~\bibnamefont
  {Bahl}}\ and\ \bibinfo {author} {\bibfnamefont {S.}~\bibnamefont {Brass}},\
  }\href {\doibase 10.1007/JHEP03(2022)017} {\bibfield  {journal} {\bibinfo
  {journal} {JHEP}\ }\textbf {\bibinfo {volume} {03}},\ \bibinfo {pages} {017}
  (\bibinfo {year} {2022})},\ \Eprint {http://arxiv.org/abs/2110.10177}
  {arXiv:2110.10177 [hep-ph]} \BibitemShut {NoStop}%
\bibitem [{\citenamefont {Arzt}\ \emph {et~al.}(1995)\citenamefont {Arzt},
  \citenamefont {Einhorn},\ and\ \citenamefont {Wudka}}]{Arzt:1994gp}%
  \BibitemOpen
  \bibfield  {author} {\bibinfo {author} {\bibfnamefont {C.}~\bibnamefont
  {Arzt}}, \bibinfo {author} {\bibfnamefont {M.~B.}\ \bibnamefont {Einhorn}}, \
  and\ \bibinfo {author} {\bibfnamefont {J.}~\bibnamefont {Wudka}},\ }\href
  {\doibase 10.1016/0550-3213(94)00336-D} {\bibfield  {journal} {\bibinfo
  {journal} {Nucl. Phys. B}\ }\textbf {\bibinfo {volume} {433}},\ \bibinfo
  {pages} {41} (\bibinfo {year} {1995})},\ \Eprint
  {http://arxiv.org/abs/hep-ph/9405214} {arXiv:hep-ph/9405214} \BibitemShut
  {NoStop}%
\bibitem [{\citenamefont {Buchalla}\ \emph {et~al.}(2023)\citenamefont
  {Buchalla}, \citenamefont {Heinrich}, \citenamefont {M\"uller-Salditt},\ and\
  \citenamefont {Pandler}}]{Buchalla:2022vjp}%
  \BibitemOpen
  \bibfield  {author} {\bibinfo {author} {\bibfnamefont {G.}~\bibnamefont
  {Buchalla}}, \bibinfo {author} {\bibfnamefont {G.}~\bibnamefont {Heinrich}},
  \bibinfo {author} {\bibfnamefont {C.}~\bibnamefont {M\"uller-Salditt}}, \
  and\ \bibinfo {author} {\bibfnamefont {F.}~\bibnamefont {Pandler}},\ }\href
  {\doibase 10.21468/SciPostPhys.15.3.088} {\bibfield  {journal} {\bibinfo
  {journal} {SciPost Phys}\ }\textbf {\bibinfo {volume} {15}},\ \bibinfo
  {pages} {088} (\bibinfo {year} {2023})},\ \Eprint
  {http://arxiv.org/abs/2204.11808} {arXiv:2204.11808 [hep-ph]} \BibitemShut
  {NoStop}%
\bibitem [{\citenamefont {Alasfar}\ \emph {et~al.}(2022)\citenamefont
  {Alasfar}, \citenamefont {de~Blas},\ and\ \citenamefont
  {Gr\"ober}}]{Alasfar:2022zyr}%
  \BibitemOpen
  \bibfield  {author} {\bibinfo {author} {\bibfnamefont {L.}~\bibnamefont
  {Alasfar}}, \bibinfo {author} {\bibfnamefont {J.}~\bibnamefont {de~Blas}}, \
  and\ \bibinfo {author} {\bibfnamefont {R.}~\bibnamefont {Gr\"ober}},\ }\href
  {\doibase 10.1007/JHEP05(2022)111} {\bibfield  {journal} {\bibinfo  {journal}
  {JHEP}\ }\textbf {\bibinfo {volume} {05}},\ \bibinfo {pages} {111} (\bibinfo
  {year} {2022})},\ \Eprint {http://arxiv.org/abs/2202.02333} {arXiv:2202.02333
  [hep-ph]} \BibitemShut {NoStop}%
\bibitem [{\citenamefont {Dawson}\ \emph {et~al.}(2003)\citenamefont {Dawson},
  \citenamefont {Jackson}, \citenamefont {Orr}, \citenamefont {Reina},\ and\
  \citenamefont {Wackeroth}}]{Dawson:2003zu}%
  \BibitemOpen
  \bibfield  {author} {\bibinfo {author} {\bibfnamefont {S.}~\bibnamefont
  {Dawson}}, \bibinfo {author} {\bibfnamefont {C.}~\bibnamefont {Jackson}},
  \bibinfo {author} {\bibfnamefont {L.~H.}\ \bibnamefont {Orr}}, \bibinfo
  {author} {\bibfnamefont {L.}~\bibnamefont {Reina}}, \ and\ \bibinfo {author}
  {\bibfnamefont {D.}~\bibnamefont {Wackeroth}},\ }\href {\doibase
  10.1103/PhysRevD.68.034022} {\bibfield  {journal} {\bibinfo  {journal} {Phys.
  Rev. D}\ }\textbf {\bibinfo {volume} {68}},\ \bibinfo {pages} {034022}
  (\bibinfo {year} {2003})},\ \Eprint {http://arxiv.org/abs/hep-ph/0305087}
  {arXiv:hep-ph/0305087} \BibitemShut {NoStop}%
\bibitem [{\citenamefont {Beenakker}\ \emph {et~al.}(2003)\citenamefont
  {Beenakker}, \citenamefont {Dittmaier}, \citenamefont {Kramer}, \citenamefont
  {Plumper}, \citenamefont {Spira},\ and\ \citenamefont
  {Zerwas}}]{Beenakker:2002nc}%
  \BibitemOpen
  \bibfield  {author} {\bibinfo {author} {\bibfnamefont {W.}~\bibnamefont
  {Beenakker}}, \bibinfo {author} {\bibfnamefont {S.}~\bibnamefont
  {Dittmaier}}, \bibinfo {author} {\bibfnamefont {M.}~\bibnamefont {Kramer}},
  \bibinfo {author} {\bibfnamefont {B.}~\bibnamefont {Plumper}}, \bibinfo
  {author} {\bibfnamefont {M.}~\bibnamefont {Spira}}, \ and\ \bibinfo {author}
  {\bibfnamefont {P.~M.}\ \bibnamefont {Zerwas}},\ }\href {\doibase
  10.1016/S0550-3213(03)00044-0} {\bibfield  {journal} {\bibinfo  {journal}
  {Nucl. Phys. B}\ }\textbf {\bibinfo {volume} {653}},\ \bibinfo {pages} {151}
  (\bibinfo {year} {2003})},\ \Eprint {http://arxiv.org/abs/hep-ph/0211352}
  {arXiv:hep-ph/0211352} \BibitemShut {NoStop}%
\bibitem [{\citenamefont {Zhang}\ \emph {et~al.}(2014)\citenamefont {Zhang},
  \citenamefont {Ma}, \citenamefont {Zhang}, \citenamefont {Chen},\ and\
  \citenamefont {Guo}}]{Zhang:2014gcy}%
  \BibitemOpen
  \bibfield  {author} {\bibinfo {author} {\bibfnamefont {Y.}~\bibnamefont
  {Zhang}}, \bibinfo {author} {\bibfnamefont {W.-G.}\ \bibnamefont {Ma}},
  \bibinfo {author} {\bibfnamefont {R.-Y.}\ \bibnamefont {Zhang}}, \bibinfo
  {author} {\bibfnamefont {C.}~\bibnamefont {Chen}}, \ and\ \bibinfo {author}
  {\bibfnamefont {L.}~\bibnamefont {Guo}},\ }\href {\doibase
  10.1016/j.physletb.2014.09.022} {\bibfield  {journal} {\bibinfo  {journal}
  {Phys. Lett. B}\ }\textbf {\bibinfo {volume} {738}},\ \bibinfo {pages} {1}
  (\bibinfo {year} {2014})},\ \Eprint {http://arxiv.org/abs/1407.1110}
  {arXiv:1407.1110 [hep-ph]} \BibitemShut {NoStop}%
\bibitem [{\citenamefont {Frixione}\ \emph {et~al.}(2014)\citenamefont
  {Frixione}, \citenamefont {Hirschi}, \citenamefont {Pagani}, \citenamefont
  {Shao},\ and\ \citenamefont {Zaro}}]{Frixione:2014qaa}%
  \BibitemOpen
  \bibfield  {author} {\bibinfo {author} {\bibfnamefont {S.}~\bibnamefont
  {Frixione}}, \bibinfo {author} {\bibfnamefont {V.}~\bibnamefont {Hirschi}},
  \bibinfo {author} {\bibfnamefont {D.}~\bibnamefont {Pagani}}, \bibinfo
  {author} {\bibfnamefont {H.~S.}\ \bibnamefont {Shao}}, \ and\ \bibinfo
  {author} {\bibfnamefont {M.}~\bibnamefont {Zaro}},\ }\href {\doibase
  10.1007/JHEP09(2014)065} {\bibfield  {journal} {\bibinfo  {journal} {JHEP}\
  }\textbf {\bibinfo {volume} {09}},\ \bibinfo {pages} {065} (\bibinfo {year}
  {2014})},\ \Eprint {http://arxiv.org/abs/1407.0823} {arXiv:1407.0823
  [hep-ph]} \BibitemShut {NoStop}%
\bibitem [{\citenamefont {Di~Noi}\ \emph {et~al.}(2023)\citenamefont {Di~Noi},
  \citenamefont {Gr\"ober}, \citenamefont {Heinrich}, \citenamefont {Lang},\
  and\ \citenamefont {Vitti}}]{DiNoi:2023ygk}%
  \BibitemOpen
  \bibfield  {author} {\bibinfo {author} {\bibfnamefont {S.}~\bibnamefont
  {Di~Noi}}, \bibinfo {author} {\bibfnamefont {R.}~\bibnamefont {Gr\"ober}},
  \bibinfo {author} {\bibfnamefont {G.}~\bibnamefont {Heinrich}}, \bibinfo
  {author} {\bibfnamefont {J.}~\bibnamefont {Lang}}, \ and\ \bibinfo {author}
  {\bibfnamefont {M.}~\bibnamefont {Vitti}},\ }\href@noop {} {\  (\bibinfo
  {year} {2023})},\ \Eprint {http://arxiv.org/abs/2310.18221} {arXiv:2310.18221
  [hep-ph]} \BibitemShut {NoStop}%
\bibitem [{\citenamefont {{'t Hooft}}\ and\ \citenamefont
  {Veltman}(1972)}]{THOOFT1972189}%
  \BibitemOpen
  \bibfield  {author} {\bibinfo {author} {\bibfnamefont {G.}~\bibnamefont {{'t
  Hooft}}}\ and\ \bibinfo {author} {\bibfnamefont {M.}~\bibnamefont
  {Veltman}},\ }\href {\doibase https://doi.org/10.1016/0550-3213(72)90279-9}
  {\bibfield  {journal} {\bibinfo  {journal} {Nuclear Physics B}\ }\textbf
  {\bibinfo {volume} {44}},\ \bibinfo {pages} {189} (\bibinfo {year}
  {1972})}\BibitemShut {NoStop}%
\bibitem [{\citenamefont {Breitenlohner}\ and\ \citenamefont
  {Maison}(1977)}]{Breitenlohner:1977hr}%
  \BibitemOpen
  \bibfield  {author} {\bibinfo {author} {\bibfnamefont {P.}~\bibnamefont
  {Breitenlohner}}\ and\ \bibinfo {author} {\bibfnamefont {D.}~\bibnamefont
  {Maison}},\ }\href {\doibase 10.1007/BF01609069} {\bibfield  {journal}
  {\bibinfo  {journal} {Commun. Math. Phys.}\ }\textbf {\bibinfo {volume}
  {52}},\ \bibinfo {pages} {11} (\bibinfo {year} {1977})}\BibitemShut {NoStop}%
\bibitem [{\citenamefont {Nogueira}(1993)}]{Nogueira:1991ex}%
  \BibitemOpen
  \bibfield  {author} {\bibinfo {author} {\bibfnamefont {P.}~\bibnamefont
  {Nogueira}},\ }\href {\doibase 10.1006/jcph.1993.1074} {\bibfield  {journal}
  {\bibinfo  {journal} {J. Comput. Phys.}\ }\textbf {\bibinfo {volume} {105}},\
  \bibinfo {pages} {279} (\bibinfo {year} {1993})}\BibitemShut {NoStop}%
\bibitem [{\citenamefont {Mertig}\ \emph {et~al.}(1991)\citenamefont {Mertig},
  \citenamefont {Bohm},\ and\ \citenamefont {Denner}}]{Mertig:1990an}%
  \BibitemOpen
  \bibfield  {author} {\bibinfo {author} {\bibfnamefont {R.}~\bibnamefont
  {Mertig}}, \bibinfo {author} {\bibfnamefont {M.}~\bibnamefont {Bohm}}, \ and\
  \bibinfo {author} {\bibfnamefont {A.}~\bibnamefont {Denner}},\ }\href
  {\doibase 10.1016/0010-4655(91)90130-D} {\bibfield  {journal} {\bibinfo
  {journal} {Comput. Phys. Commun.}\ }\textbf {\bibinfo {volume} {64}},\
  \bibinfo {pages} {345} (\bibinfo {year} {1991})}\BibitemShut {NoStop}%
\bibitem [{\citenamefont {Shtabovenko}\ \emph {et~al.}(2016)\citenamefont
  {Shtabovenko}, \citenamefont {Mertig},\ and\ \citenamefont
  {Orellana}}]{Shtabovenko:2016sxi}%
  \BibitemOpen
  \bibfield  {author} {\bibinfo {author} {\bibfnamefont {V.}~\bibnamefont
  {Shtabovenko}}, \bibinfo {author} {\bibfnamefont {R.}~\bibnamefont {Mertig}},
  \ and\ \bibinfo {author} {\bibfnamefont {F.}~\bibnamefont {Orellana}},\
  }\href {\doibase 10.1016/j.cpc.2016.06.008} {\bibfield  {journal} {\bibinfo
  {journal} {Comput. Phys. Commun.}\ }\textbf {\bibinfo {volume} {207}},\
  \bibinfo {pages} {432} (\bibinfo {year} {2016})},\ \Eprint
  {http://arxiv.org/abs/1601.01167} {arXiv:1601.01167 [hep-ph]} \BibitemShut
  {NoStop}%
\bibitem [{\citenamefont {Shtabovenko}\ \emph {et~al.}(2020)\citenamefont
  {Shtabovenko}, \citenamefont {Mertig},\ and\ \citenamefont
  {Orellana}}]{Shtabovenko:2020gxv}%
  \BibitemOpen
  \bibfield  {author} {\bibinfo {author} {\bibfnamefont {V.}~\bibnamefont
  {Shtabovenko}}, \bibinfo {author} {\bibfnamefont {R.}~\bibnamefont {Mertig}},
  \ and\ \bibinfo {author} {\bibfnamefont {F.}~\bibnamefont {Orellana}},\
  }\href {\doibase 10.1016/j.cpc.2020.107478} {\bibfield  {journal} {\bibinfo
  {journal} {Comput. Phys. Commun.}\ }\textbf {\bibinfo {volume} {256}},\
  \bibinfo {pages} {107478} (\bibinfo {year} {2020})},\ \Eprint
  {http://arxiv.org/abs/2001.04407} {arXiv:2001.04407 [hep-ph]} \BibitemShut
  {NoStop}%
\bibitem [{\citenamefont {Kachelriess}\ and\ \citenamefont
  {Malmquist}(2022)}]{externalgluons}%
  \BibitemOpen
  \bibfield  {author} {\bibinfo {author} {\bibfnamefont {M.}~\bibnamefont
  {Kachelriess}}\ and\ \bibinfo {author} {\bibfnamefont {M.~N.}\ \bibnamefont
  {Malmquist}},\ }\href {\doibase 10.1140/epjp/s13360-021-02273-3} {\bibfield
  {journal} {\bibinfo  {journal} {Eur. Phys. J. Plus}\ }\textbf {\bibinfo
  {volume} {137}},\ \bibinfo {pages} {89} (\bibinfo {year} {2022})},\ \Eprint
  {http://arxiv.org/abs/2107.07187} {arXiv:2107.07187 [hep-ph]} \BibitemShut
  {NoStop}%
\bibitem [{\citenamefont {Kleiss}\ \emph {et~al.}(1986)\citenamefont {Kleiss},
  \citenamefont {Stirling},\ and\ \citenamefont {Ellis}}]{rambo}%
  \BibitemOpen
  \bibfield  {author} {\bibinfo {author} {\bibfnamefont {R.~H.}\ \bibnamefont
  {Kleiss}}, \bibinfo {author} {\bibfnamefont {W.~J.}\ \bibnamefont
  {Stirling}}, \ and\ \bibinfo {author} {\bibfnamefont {S.~D.}\ \bibnamefont
  {Ellis}},\ }\href {\doibase 10.1016/0010-4655(86)90119-0} {\bibfield
  {journal} {\bibinfo  {journal} {Comput. Phys. Commun.}\ }\textbf {\bibinfo
  {volume} {40}},\ \bibinfo {pages} {359} (\bibinfo {year} {1986})}\BibitemShut
  {NoStop}%
\bibitem [{\citenamefont {Buckley}\ \emph {et~al.}(2015)\citenamefont
  {Buckley}, \citenamefont {Ferrando}, \citenamefont {Lloyd}, \citenamefont
  {Nordstr\"om}, \citenamefont {Page}, \citenamefont {R\"ufenacht},
  \citenamefont {Sch\"onherr},\ and\ \citenamefont {Watt}}]{lhapdf}%
  \BibitemOpen
  \bibfield  {author} {\bibinfo {author} {\bibfnamefont {A.}~\bibnamefont
  {Buckley}}, \bibinfo {author} {\bibfnamefont {J.}~\bibnamefont {Ferrando}},
  \bibinfo {author} {\bibfnamefont {S.}~\bibnamefont {Lloyd}}, \bibinfo
  {author} {\bibfnamefont {K.}~\bibnamefont {Nordstr\"om}}, \bibinfo {author}
  {\bibfnamefont {B.}~\bibnamefont {Page}}, \bibinfo {author} {\bibfnamefont
  {M.}~\bibnamefont {R\"ufenacht}}, \bibinfo {author} {\bibfnamefont
  {M.}~\bibnamefont {Sch\"onherr}}, \ and\ \bibinfo {author} {\bibfnamefont
  {G.}~\bibnamefont {Watt}},\ }\href {\doibase 10.1140/epjc/s10052-015-3318-8}
  {\bibfield  {journal} {\bibinfo  {journal} {Eur. Phys. J. C}\ }\textbf
  {\bibinfo {volume} {75}},\ \bibinfo {pages} {132} (\bibinfo {year} {2015})},\
  \Eprint {http://arxiv.org/abs/1412.7420} {arXiv:1412.7420 [hep-ph]}
  \BibitemShut {NoStop}%
\bibitem [{\citenamefont {Ball}\ \emph {et~al.}(2022)\citenamefont {Ball} \emph
  {et~al.}}]{NNPDF:2021njg}%
  \BibitemOpen
  \bibfield  {author} {\bibinfo {author} {\bibfnamefont {R.~D.}\ \bibnamefont
  {Ball}} \emph {et~al.} (\bibinfo {collaboration} {NNPDF}),\ }\href {\doibase
  10.1140/epjc/s10052-022-10328-7} {\bibfield  {journal} {\bibinfo  {journal}
  {Eur. Phys. J. C}\ }\textbf {\bibinfo {volume} {82}},\ \bibinfo {pages} {428}
  (\bibinfo {year} {2022})},\ \Eprint {http://arxiv.org/abs/2109.02653}
  {arXiv:2109.02653 [hep-ph]} \BibitemShut {NoStop}%
\bibitem [{\citenamefont {Di~Noi}\ and\ \citenamefont
  {Silvestrini}(2023)}]{rgesolver}%
  \BibitemOpen
  \bibfield  {author} {\bibinfo {author} {\bibfnamefont {S.}~\bibnamefont
  {Di~Noi}}\ and\ \bibinfo {author} {\bibfnamefont {L.}~\bibnamefont
  {Silvestrini}},\ }\href {\doibase 10.1140/epjc/s10052-023-11189-4} {\bibfield
   {journal} {\bibinfo  {journal} {Eur. Phys. J. C}\ }\textbf {\bibinfo
  {volume} {83}},\ \bibinfo {pages} {200} (\bibinfo {year} {2023})},\ \Eprint
  {http://arxiv.org/abs/2210.06838} {arXiv:2210.06838 [hep-ph]} \BibitemShut
  {NoStop}%
\bibitem [{\citenamefont {Fuentes-Martin}\ \emph {et~al.}(2021)\citenamefont
  {Fuentes-Martin}, \citenamefont {Ruiz-Femenia}, \citenamefont {Vicente},\
  and\ \citenamefont {Virto}}]{Fuentes-Martin:2020zaz}%
  \BibitemOpen
  \bibfield  {author} {\bibinfo {author} {\bibfnamefont {J.}~\bibnamefont
  {Fuentes-Martin}}, \bibinfo {author} {\bibfnamefont {P.}~\bibnamefont
  {Ruiz-Femenia}}, \bibinfo {author} {\bibfnamefont {A.}~\bibnamefont
  {Vicente}}, \ and\ \bibinfo {author} {\bibfnamefont {J.}~\bibnamefont
  {Virto}},\ }\href {\doibase 10.1140/epjc/s10052-020-08778-y} {\bibfield
  {journal} {\bibinfo  {journal} {Eur. Phys. J. C}\ }\textbf {\bibinfo {volume}
  {81}},\ \bibinfo {pages} {167} (\bibinfo {year} {2021})},\ \Eprint
  {http://arxiv.org/abs/2010.16341} {arXiv:2010.16341 [hep-ph]} \BibitemShut
  {NoStop}%
\bibitem [{\citenamefont {Aebischer}\ \emph {et~al.}(2018)\citenamefont
  {Aebischer}, \citenamefont {Kumar},\ and\ \citenamefont
  {Straub}}]{Aebischer:2018bkb}%
  \BibitemOpen
  \bibfield  {author} {\bibinfo {author} {\bibfnamefont {J.}~\bibnamefont
  {Aebischer}}, \bibinfo {author} {\bibfnamefont {J.}~\bibnamefont {Kumar}}, \
  and\ \bibinfo {author} {\bibfnamefont {D.~M.}\ \bibnamefont {Straub}},\
  }\href {\doibase 10.1140/epjc/s10052-018-6492-7} {\bibfield  {journal}
  {\bibinfo  {journal} {Eur. Phys. J. C}\ }\textbf {\bibinfo {volume} {78}},\
  \bibinfo {pages} {1026} (\bibinfo {year} {2018})},\ \Eprint
  {http://arxiv.org/abs/1804.05033} {arXiv:1804.05033 [hep-ph]} \BibitemShut
  {NoStop}%
\bibitem [{\citenamefont {Brivio}(2021)}]{Brivio:2020onw}%
  \BibitemOpen
  \bibfield  {author} {\bibinfo {author} {\bibfnamefont {I.}~\bibnamefont
  {Brivio}},\ }\href {\doibase 10.1007/JHEP04(2021)073} {\bibfield  {journal}
  {\bibinfo  {journal} {JHEP}\ }\textbf {\bibinfo {volume} {04}},\ \bibinfo
  {pages} {073} (\bibinfo {year} {2021})},\ \Eprint
  {http://arxiv.org/abs/2012.11343} {arXiv:2012.11343 [hep-ph]} \BibitemShut
  {NoStop}%
\bibitem [{\citenamefont {Corbett}\ \emph {et~al.}(2021)\citenamefont
  {Corbett}, \citenamefont {Martin},\ and\ \citenamefont
  {Trott}}]{Corbett:2021cil}%
  \BibitemOpen
  \bibfield  {author} {\bibinfo {author} {\bibfnamefont {T.}~\bibnamefont
  {Corbett}}, \bibinfo {author} {\bibfnamefont {A.}~\bibnamefont {Martin}}, \
  and\ \bibinfo {author} {\bibfnamefont {M.}~\bibnamefont {Trott}},\ }\href
  {\doibase 10.1007/JHEP12(2021)147} {\bibfield  {journal} {\bibinfo  {journal}
  {JHEP}\ }\textbf {\bibinfo {volume} {12}},\ \bibinfo {pages} {147} (\bibinfo
  {year} {2021})},\ \Eprint {http://arxiv.org/abs/2107.07470} {arXiv:2107.07470
  [hep-ph]} \BibitemShut {NoStop}%
\bibitem [{\citenamefont {Martin}\ and\ \citenamefont
  {Trott}(2023)}]{Martin:2023fad}%
  \BibitemOpen
  \bibfield  {author} {\bibinfo {author} {\bibfnamefont {A.}~\bibnamefont
  {Martin}}\ and\ \bibinfo {author} {\bibfnamefont {M.}~\bibnamefont {Trott}},\
  }\href@noop {} {\  (\bibinfo {year} {2023})},\ \Eprint
  {http://arxiv.org/abs/2305.05879} {arXiv:2305.05879 [hep-ph]} \BibitemShut
  {NoStop}%
\bibitem [{\citenamefont {Biek\"otter}\ \emph
  {et~al.}(2023{\natexlab{a}})\citenamefont {Biek\"otter}, \citenamefont
  {Pecjak}, \citenamefont {Scott},\ and\ \citenamefont
  {Smith}}]{Biekotter:2023xle}%
  \BibitemOpen
  \bibfield  {author} {\bibinfo {author} {\bibfnamefont {A.}~\bibnamefont
  {Biek\"otter}}, \bibinfo {author} {\bibfnamefont {B.~D.}\ \bibnamefont
  {Pecjak}}, \bibinfo {author} {\bibfnamefont {D.~J.}\ \bibnamefont {Scott}}, \
  and\ \bibinfo {author} {\bibfnamefont {T.}~\bibnamefont {Smith}},\ }\href
  {\doibase 10.1007/JHEP07(2023)115} {\bibfield  {journal} {\bibinfo  {journal}
  {JHEP}\ }\textbf {\bibinfo {volume} {07}},\ \bibinfo {pages} {115} (\bibinfo
  {year} {2023}{\natexlab{a}})},\ \Eprint {http://arxiv.org/abs/2305.03763}
  {arXiv:2305.03763 [hep-ph]} \BibitemShut {NoStop}%
\bibitem [{\citenamefont {Biek\"otter}\ \emph
  {et~al.}(2023{\natexlab{b}})\citenamefont {Biek\"otter}, \citenamefont
  {Pecjak},\ and\ \citenamefont {Smith}}]{Biekotter:2023xxx}%
  \BibitemOpen
  \bibfield  {author} {\bibinfo {author} {\bibfnamefont {A.}~\bibnamefont
  {Biek\"otter}}, \bibinfo {author} {\bibfnamefont {B.~D.}\ \bibnamefont
  {Pecjak}}, \ and\ \bibinfo {author} {\bibfnamefont {T.}~\bibnamefont
  {Smith}},\ }\href@noop {} {\  (\bibinfo {year} {2023}{\natexlab{b}})},\
  \Eprint {http://arxiv.org/abs/2312.08446} {arXiv:2312.08446 [hep-ph]}
  \BibitemShut {NoStop}%
\bibitem [{\citenamefont {de~Blas}\ \emph {et~al.}(2018)\citenamefont
  {de~Blas}, \citenamefont {Criado}, \citenamefont {Perez-Victoria},\ and\
  \citenamefont {Santiago}}]{deBlas:2017xtg}%
  \BibitemOpen
  \bibfield  {author} {\bibinfo {author} {\bibfnamefont {J.}~\bibnamefont
  {de~Blas}}, \bibinfo {author} {\bibfnamefont {J.~C.}\ \bibnamefont {Criado}},
  \bibinfo {author} {\bibfnamefont {M.}~\bibnamefont {Perez-Victoria}}, \ and\
  \bibinfo {author} {\bibfnamefont {J.}~\bibnamefont {Santiago}},\ }\href
  {\doibase 10.1007/JHEP03(2018)109} {\bibfield  {journal} {\bibinfo  {journal}
  {JHEP}\ }\textbf {\bibinfo {volume} {03}},\ \bibinfo {pages} {109} (\bibinfo
  {year} {2018})},\ \Eprint {http://arxiv.org/abs/1711.10391} {arXiv:1711.10391
  [hep-ph]} \BibitemShut {NoStop}%
\bibitem [{\citenamefont {Deutschmann}\ \emph {et~al.}(2017)\citenamefont
  {Deutschmann}, \citenamefont {Duhr}, \citenamefont {Maltoni},\ and\
  \citenamefont {Vryonidou}}]{Deutschmann:2017qum}%
  \BibitemOpen
  \bibfield  {author} {\bibinfo {author} {\bibfnamefont {N.}~\bibnamefont
  {Deutschmann}}, \bibinfo {author} {\bibfnamefont {C.}~\bibnamefont {Duhr}},
  \bibinfo {author} {\bibfnamefont {F.}~\bibnamefont {Maltoni}}, \ and\
  \bibinfo {author} {\bibfnamefont {E.}~\bibnamefont {Vryonidou}},\ }\href
  {\doibase 10.1007/JHEP12(2017)063} {\bibfield  {journal} {\bibinfo  {journal}
  {JHEP}\ }\textbf {\bibinfo {volume} {12}},\ \bibinfo {pages} {063} (\bibinfo
  {year} {2017})},\ \bibinfo {note} {[Erratum: JHEP 02, 159 (2018)]},\ \Eprint
  {http://arxiv.org/abs/1708.00460} {arXiv:1708.00460 [hep-ph]} \BibitemShut
  {NoStop}%
\bibitem [{\citenamefont {Jenkins}\ \emph {et~al.}(2023)\citenamefont
  {Jenkins}, \citenamefont {Manohar}, \citenamefont {Naterop},\ and\
  \citenamefont {Pag\`es}}]{Jenkins:2023bls}%
  \BibitemOpen
  \bibfield  {author} {\bibinfo {author} {\bibfnamefont {E.~E.}\ \bibnamefont
  {Jenkins}}, \bibinfo {author} {\bibfnamefont {A.~V.}\ \bibnamefont
  {Manohar}}, \bibinfo {author} {\bibfnamefont {L.}~\bibnamefont {Naterop}}, \
  and\ \bibinfo {author} {\bibfnamefont {J.}~\bibnamefont {Pag\`es}},\
  }\href@noop {} {\  (\bibinfo {year} {2023})},\ \Eprint
  {http://arxiv.org/abs/2310.19883} {arXiv:2310.19883 [hep-ph]} \BibitemShut
  {NoStop}%
\bibitem [{\citenamefont {Bern}\ \emph {et~al.}(2020)\citenamefont {Bern},
  \citenamefont {Parra-Martinez},\ and\ \citenamefont {Sawyer}}]{Bern:2020ikv}%
  \BibitemOpen
  \bibfield  {author} {\bibinfo {author} {\bibfnamefont {Z.}~\bibnamefont
  {Bern}}, \bibinfo {author} {\bibfnamefont {J.}~\bibnamefont
  {Parra-Martinez}}, \ and\ \bibinfo {author} {\bibfnamefont {E.}~\bibnamefont
  {Sawyer}},\ }\href {\doibase 10.1007/JHEP10(2020)211} {\bibfield  {journal}
  {\bibinfo  {journal} {JHEP}\ }\textbf {\bibinfo {volume} {10}},\ \bibinfo
  {pages} {211} (\bibinfo {year} {2020})},\ \Eprint
  {http://arxiv.org/abs/2005.12917} {arXiv:2005.12917 [hep-ph]} \BibitemShut
  {NoStop}%
\bibitem [{\citenamefont {Ellis}(2017)}]{Ellis:2016jkw}%
  \BibitemOpen
  \bibfield  {author} {\bibinfo {author} {\bibfnamefont {J.}~\bibnamefont
  {Ellis}},\ }\href {\doibase 10.1016/j.cpc.2016.08.019} {\bibfield  {journal}
  {\bibinfo  {journal} {Comput. Phys. Commun.}\ }\textbf {\bibinfo {volume}
  {210}},\ \bibinfo {pages} {103} (\bibinfo {year} {2017})},\ \Eprint
  {http://arxiv.org/abs/1601.05437} {arXiv:1601.05437 [hep-ph]} \BibitemShut
  {NoStop}%
\end{thebibliography}%

\end{document}